\documentclass{iau-JDSS}
\usepackage{graphicx}
\usepackage{psfig}

\title[On the present and future of pulsar astronomy]
{On the present and future of pulsar astronomy}

\author[Becker, Gil \& Rudak 2007]
{W. Becker$^1$ \and J.~Gil,$^2$ \and B.~Rudak$^3$}

\affiliation{
$^1$ Max-Planck Institut f\"ur extraterr.~Physik, Giessenbachstr.~1, 85741 Garching, Germany,
\break email: web@mpe.mpg.de\\[\affilskip]
$^2$ J.~Kepler Institute of Astronomy, University of Zielona Gora, Poland,
\break email: jag@astro.ia.uz.zgora.pl\\[\affilskip]
$^3$CAMK, Rabianska 8, 87-100 Torun, Poland\break email: bronek@ncac.torun.pl}

\pubyear{2007}
\volume{Volume 14}
\pagerange{1--30}
\date{?? and in revised form ??}
\jname{Highlights of Astronomy, Volume 14}
\editors{K.A. van der Hucht, ed.}

\begin{document}
\maketitle

\section{Introduction}

\noindent
Neutron stars are formed in supernova explosions. They manifest themselves in
many different ways, for example, as pulsars, anomalous X-ray pulsars (AXPs)
and soft gamma-ray repeaters (SGRs) and the so-called 'radio-quiet neutron
stars'. These objects are made visible by high-energy processes occurring
on their surface or in the surrounding region. In most of these objects,
ultra-strong magnetic fields are a crucial element in the radio, optical,
X-ray and gamma-ray emission processes which dominate the observed spectrum.

The physics of pulsars spans a wide range of disciplines, from nuclear and
condensed matter physics of very dense matter in neutron star interiors, to
plasma physics and electrodynamics of the magnetospheres, to relativistic
magnetohydrodynamics of electron-positron pulsar winds interacting with the
surrounding ambient medium. Not to forget the test bed they provide for
general relativity theories as well as being sources of gravitational waves.

Observationally, pulsar research is advancing rapidly. 
A great array of space instruments (the Hubble Space Telescope, ROSAT, ASCA, 
BeppoSAX, RXTE and the Compton Gamma-Ray Observatory), launched in the last 
decade of the previous century, have opened new windows in pulsar research 
with high quality data in energy bands from optical to the gamma-rays. With 
the more recently launched satellite X-ray observatories CHANDRA and XMM-Newton, 
upgraded radio observatories and ground based optical telescopes like Keck, GEMINI,
Subaru and the VLT lots of the questions remained unanswered from the previous
generation of observatories could be addressed and resulted in new and exciting
results which changed the previous picture of neutron star evolution substantially
(e.g. making Crab-like pulsars the exception not the rule for the appearance of
a young neutron star). The X-ray Observatory SUZAKU (formerly ASTRO-E2) launched 
in 2005, and the gamma-ray Observatories AGILE and GLAST, which are supposed to be 
launched end of 2007/beginning of 2008, will complement the pulsar studies and 
will enlarge the class of pulsars detected at their energy bands.

However, even in view of these great observational capabilities, the physical
processes responsible for the pulsars' broad-band emission, observed from the
infrared to the gamma-ray band are still poorly known. Although no uniquely
accepted theory exists so far, a notable progress was made very recently in
this respect. New developments include caustic slot gaps as well as modified
outer gaps. Last but not least: polarization-dependent treatment of high-energy
radiative processes is under way, since X-ray and gamma-ray polarimetry - a new
powerful tool to discriminate between competing models - is around the corner.

To face recent observational results obtained in multi-wavelength studies from
neutron stars and pulsars with the various theoretical models and to discuss on
future perspectives on neutron star astronomy we organized a Joined Discussion (JD02)
during the XXVI IAU General Assembly which took place in 2006 August in Prague.
More than 150 scientists took actively part in this Joint Discussion. Fourteen invited
review talks were presented to view the present and future of pulsar astronomy.
Fifty three poster contributions displayed new and exciting results. PDF-files of all
review talks and most of the posters displayed during the meeting are available
on the  meeting website {\bf http://www.mpe.mpg.de/IAU\_JD02}. 

The following sections give an overview of the invited review talks and contributed
posters. The review talks are subject of review articles which will be published
elsewhere. More information on this will be available at the URL given above.

\section{Invited Review Papers}\bigskip

\subsection{Radio emission properties of pulsars}
{\bf R.N.~Manchester:}  Currently, 1765 pulsars are known.
170 of them are millisecond pulsars. 131 pulsars are in binary systems.
129 pulsars are detected in 24 globular clusters. Recent observational
results on the radio emission properties of pulsars were reviewed.

\subsection{ New results on rotating radio transients}\smallskip
{\bf M.A.~McLaughlin:} A new population of radio-bursting neutron stars discovered
in a large scale search for transient radio sources was discussed. Unlike normal 
radio pulsars, these objects, which are called Rotating RAdio Transient (or RRATs), 
cannot be detected through their time-averaged emission and are radio sources for 
typically less than 1 second per day. The spin periods of these objects range from 
0.4 to 7 seconds, with period derivatives indicating that at least one RRAT has a 
magnetar-strength magnetic field. Recent developments were detailed, including X-ray 
observations and observations with more sensitive radio telescopes, and it was 
discussed how these objects are related to other neutron star populations. Also, 
the implications of this new source class for neutron star population estimates was 
described.

\subsection{Isolated neutron stars in optical and X-rays: room for discovery}\smallskip
{\bf P.A.~Caraveo:} The multi-wavelength behavior of isolated neutron stars evolves as they age.
In particular, the X-ray and optical emissions allow us to follow the shift from the non-thermal
regime, typical of young objects, to a mostly thermal one, typical of older specimen. New observations
unveil tale-telling details both on young and old objects, reminding us that a lot remains to be
discovered in the complex INS family tree.

\subsection{Gamma-ray and TeV emission properties of pulsars and pulsar wind nebulae}\smallskip
{\bf O.C.~De Jager:} Although more than 1,600 radio pulsars have been discovered, only a few have
been detected in the gamma-ray band. This is not because they are intrinsically faint, but because
the pulsed component seems to cut off below about 30 GeV (the EGRET range), where the sensitivity was
severely limited. However, ground-based atmospheric Cerenkov telescopes operating above 100 GeV (the
Very High Energy or VHE domain), have both good sensitivity and good angular resolution to resolve
several pulsar wind nebulae (PWN) in the VHE gamma-ray domain.
This review talk summarized the progress to date on pulsar and pulsar wind nebula observations 
and theory. Since gamma-ray observations below 30 GeV have been limited by poor
sensitivity, an instrument like GLAST should be able to resolve the pulsed component of a significant
fraction of radio pulsars. This talk showed how the discovery potential of GLAST would be limited for
fainter sources in the absence of contemporary radio pulsar parameters. This calls for the introduction
of wide field-of-view radio pulsar monitors like KAT to resolve this problem. Most progress on PWN in
the gamma-ray domain is made by the HESS telescope system in Namibia. In this case we progressed to
the level where VHE Gamma-ray Astronomy is taking the lead at all wavelengths (radio, IR, optical,
X-ray and gamma-ray) in the identification and understanding of new PWN.
It was shown how the spin history of the PWN is more relevant to such VHE observations rather 
than X-rays, although the latter probe the more recent history of PWN evolution.
It was then shown how these complementary wavebands can be combined to obtain new information 
about aspects such as the birth periods of pulsars and conversion efficiency of spin-down power 
to injected ultra-relativistic electrons. 

 \begin{figure*}
  \centerline{\psfig{figure=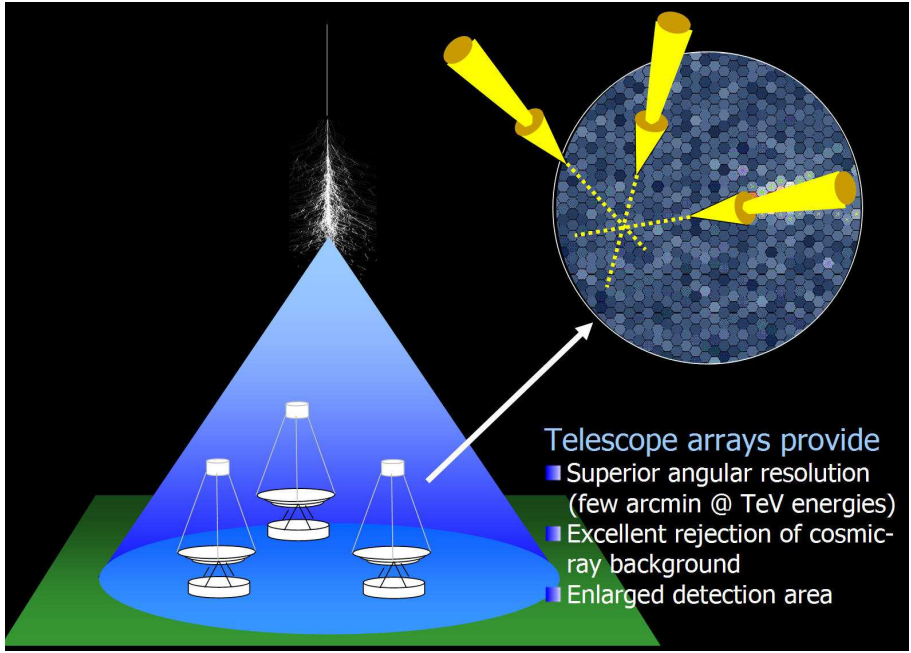,width=12cm,clip=}}
  \caption[]{\small The detection of high energy gamma rays with the e.g.~H.E.S.S. 
   telescopes is based on the imaging air Cherenkov technique. An incident
   high-energy gamma ray interacts high up in the atmosphere and generates an
   air shower of secondary particles. The number of shower particles reaches a
   maximum at about 10 km height, and the shower dies out deeper in the
   atmosphere. Since the shower particles move at essentially the speed of
   light, they emit the so-called Cherenkov light, a faint blue light. 
   The Cherenkov light is beamed around the direction of the incident primary
  particle and illuminates on the ground an area of about 250 m diameter.
  Chart from the talk presented by O.C.~De Jager.} 
  \end{figure*}

\subsection{Pulsar timing and its future perspective}\smallskip
{\bf D.J.~Nice:} The present state of radio pulsar timing and prospects for future
progress was surveyed.  In recent years, pulsar timing experiments have grown rapidly in both 
quantity and quality, yielding new tests of gravitation, new constraints on the structure of neutron 
stars, new insight into neutron star kicks and the dynamics of supernovae, and better understanding 
of the evolution of compact objects in the Galaxy.

The tremendous success of recent pulsar surveys, especially those with the Parkes multi-beam receiver,
have vastly increased the number of known pulsars.  Ongoing surveys at Arecibo and elsewhere promise
to continue yielding new and exciting pulsars.  At the same time, advances in instrumentation-- including
the development of wide-band receivers and spectrometers and the routine use of software coherent
de-dispersion data acquisition systems-- are increasing the precision attainable in timing experiments.
State of the art timing precision is now around 100 nanoseconds. Limitations have been discussed on 
present timing experiments and prospects for future improvements.

  \begin{figure*}
  \centerline{\psfig{figure=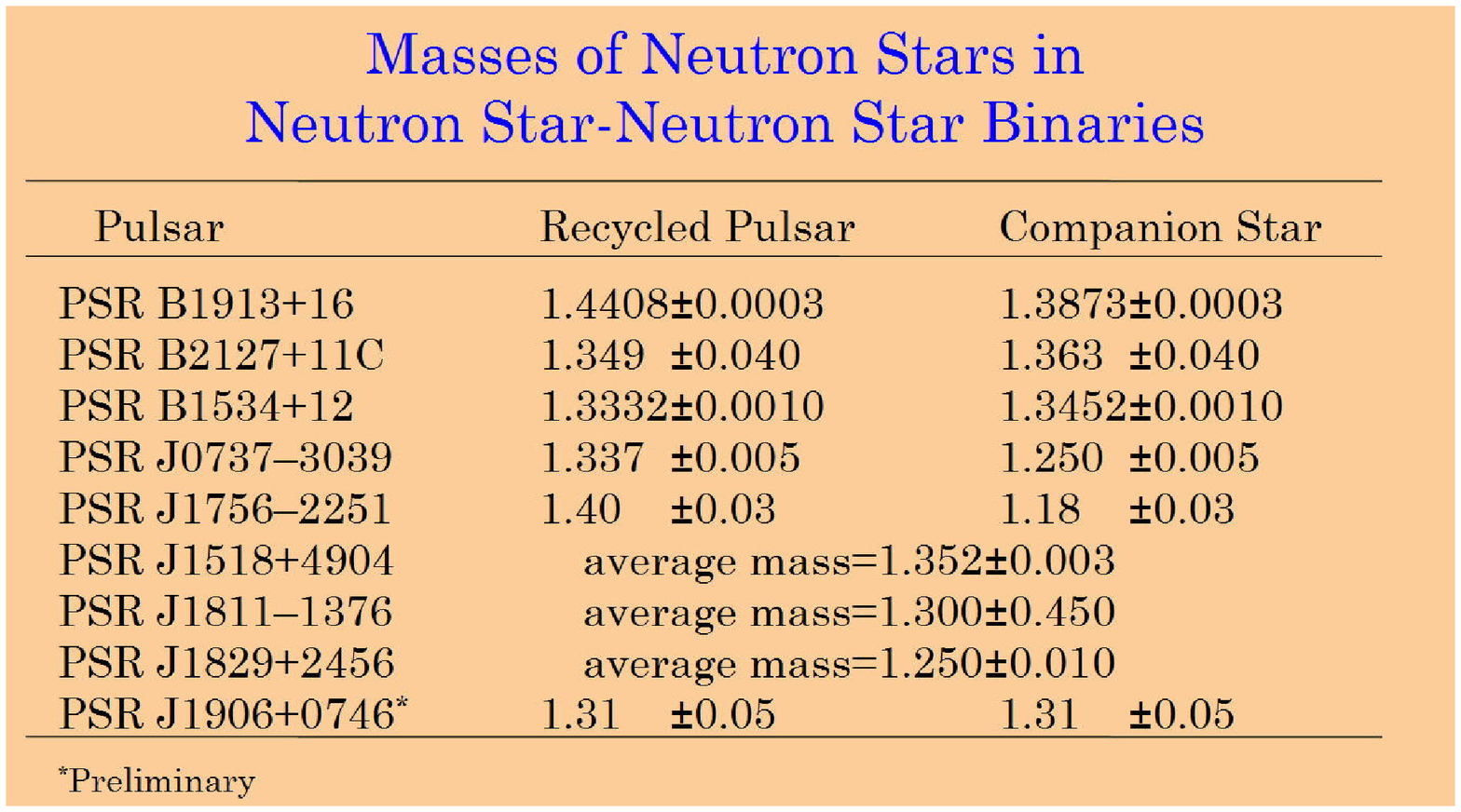,width=13cm,clip=}}
  \caption[]{\small Masses of Neutron Stars in NS-NS binaries. 
   Chart from the talk of D.~Nice.}
  \end{figure*}

\subsection{Radio emission theories of pulsars}\smallskip
{\bf V.~Usov:} Pulsar magnetospheres contain a multi-component, strongly magnetized, relativistic
plasma. The present review was mainly concerned with generation and propagation of coherent radio
emission in this plasma, emphasizing reasons why up to now there is no commonly-accepted model
of the radio emission of pulsars. Possible progress in our knowledge about the mechanism of the
pulsar radio emission was discussed.

\subsection{Theory of high energy emission from pulsars}\smallskip
{\bf K.S.~Cheng:} In this talk various models of high energy emission from pulsars were briefly reviewed.
In particular it was pointed out that the light curves can provide important constraints in the radiation
emission regions and the location of the accelerators (gaps). Furthermore, the energy dependent light
curves and phase-dependent spectrum could not be explained in terms of simple two dimensional models,
three dimensional models must be used to explain the full detail of the observed data.
A three dimensional outer gap model was presented to study the magnetospheric geometry, the light curve 
and the phase-resolved spectra of the Crab pulsar. Using a synchrotron self-Compton mechanism, the 
phase-resolved spectra with the energy range from 100 eV to 10 GeV of the Crab pulsar can also be 
explained. The observed polarization angle swing of optical photons was also used to determine 
the viewing angle.

  \begin{figure*}
  \centerline{\psfig{figure=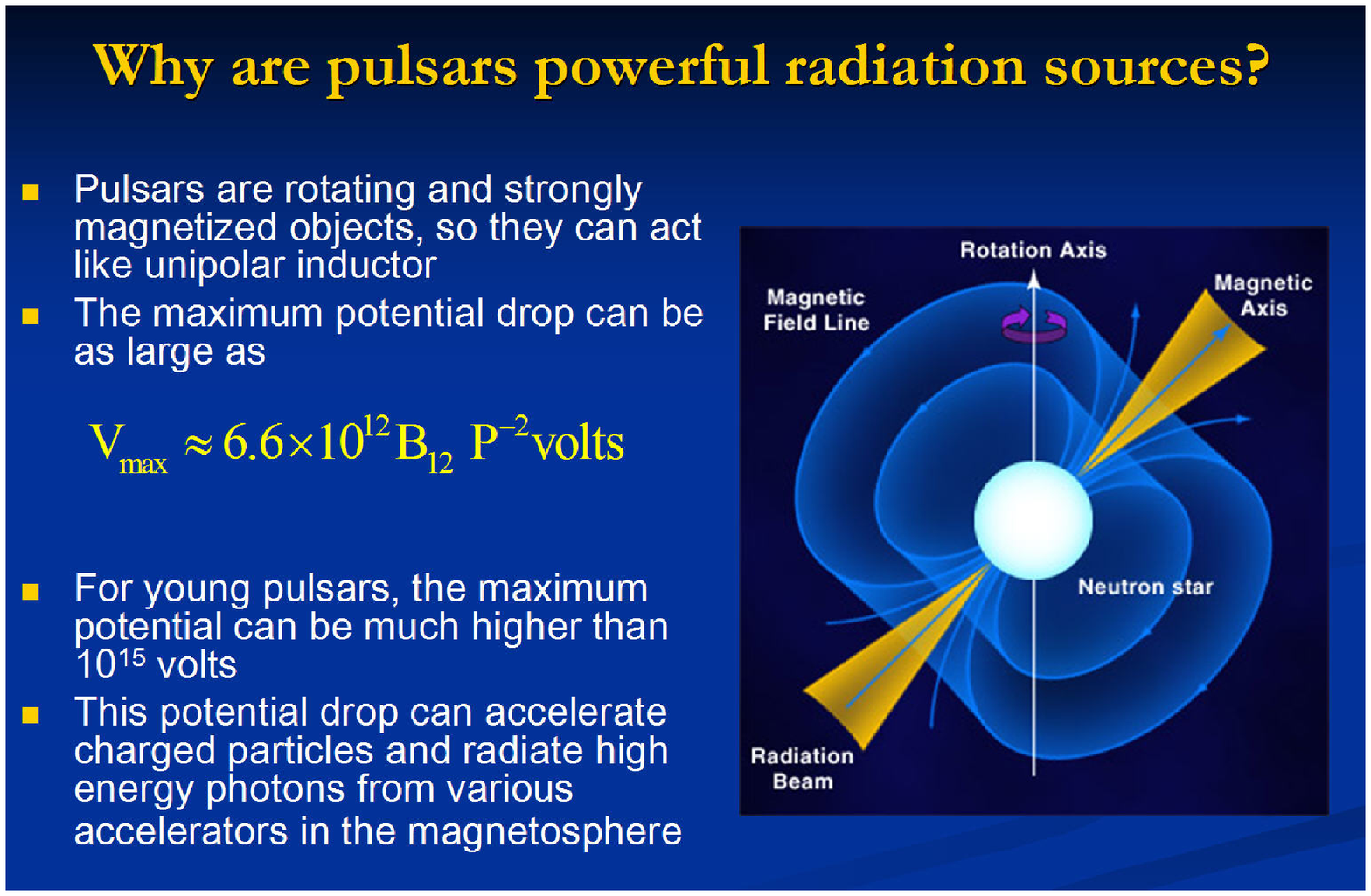,width=13cm,clip=}}
   \caption[]{\small Chart of K.S.~Cheng on why pulsars are powerful emitters of high energy radiation.}
   \end{figure*}

\subsection{On the theory of radio and high energy emission from pulsars: Where to go?}\smallskip
{\bf J.~Arons:} Recent progress in the theory of the energy loss from pulsars was discussed, focusing
on the advances in force free models of the magnetosphere and on dissipation in the wind, with 
implications of the latter for unusually models of pulsed emission.  Attention was also drawn to 
the implications the new global magnetosphere models may have for the generation of parallel electric 
fields, and the implications of these for high energy emission, radio emission and the global mass 
loss rate through pair creation. Parallel electric field formation and consequences for photon 
emission was discussed in scenarios with self-consistent currents, which are rather different from 
the standard gap models with their starvation electric fields.  Finally, remarks were presented on 
the importance of radiation transfer effects in unraveling the continuing mystery of pulsar radio 
emission.

\subsection{Cooling neutron stars: The present and the future}\smallskip
{\bf S.~Tsuruta:} Recent years have seen some significant progress in theoretical studies of physics
of dense matter. Combined with the observational data now available from the successful launch of
Chandra and XMM/Newton X-ray space missions as well as various lower-energy band observations, these
developments now offer the hope for distinguishing various competing neutron star thermal evolution
models. For instance, the latest theoretical and observational developments may already exclude both
nucleon and kaon direct Urca cooling.  In this way we can now have a realistic hope for determining
various important properties, such as the composition, superfluidity, the equation of state and
stellar radius. These developments should help us obtain deeper insight into the properties of dense
matter.

\subsection{A decade of surprises from the anomalous X-ray pulsars}\smallskip
{\bf S.M.~Ransom:}  A decade ago, the defining characteristics of the Anomalous X-ray Pulsars (AXPs)
included slow spin periods (5-9 s), relatively soft but constant X-ray luminosities in the range of
$10^{35}-10^{36}$ erg/s, and steady spin-down rates. The X-ray luminosities are too large to be 
powered by pulsar spin-down, and given the lack of evidence for accretion, are thought to be caused 
by the decay of $\sim 10^{14}-10^{15}$ G magnetic fields (i.e.~the magnetar theory as proposed by 
Thompson and Duncan). Within the past decade, though, detailed X-ray monitoring observations have 
shown that these sources are anything but constant and steady.  Timing noise, glitches, X-ray bursts, 
and pulse profile and pulsed flux variations are now known to be relatively common in these sources.  
In addition, at least one recently discovered AXP, XTE J1810-197, is a full-fledged transient object. 
Detections in the optical and infra-red (including pulsations) and recent hard X-ray observations 
have complicated our views of their emission mechanisms. Finally, very recent detections of 
magnetar-like radio pulsars, as well as strong (and transient) radio pulsations from  XTE J1810-197, 
show that these sources are linked (at least in some way) with the much more common radio pulsars.

\subsection{Pulsars and gravity}\smallskip
{\bf I.H.~Stairs:} Radio pulsars are superb tools for testing the predictions of strong-field
gravitational theories. The currently achieved tests of equivalence principles
and of predictions for relativistic binary orbital parameters were described, as well as limits on a
gravitational-wave background, and future prospects in each of these areas were discussed.
 
  \begin{figure*} 
  \centerline{\psfig{figure=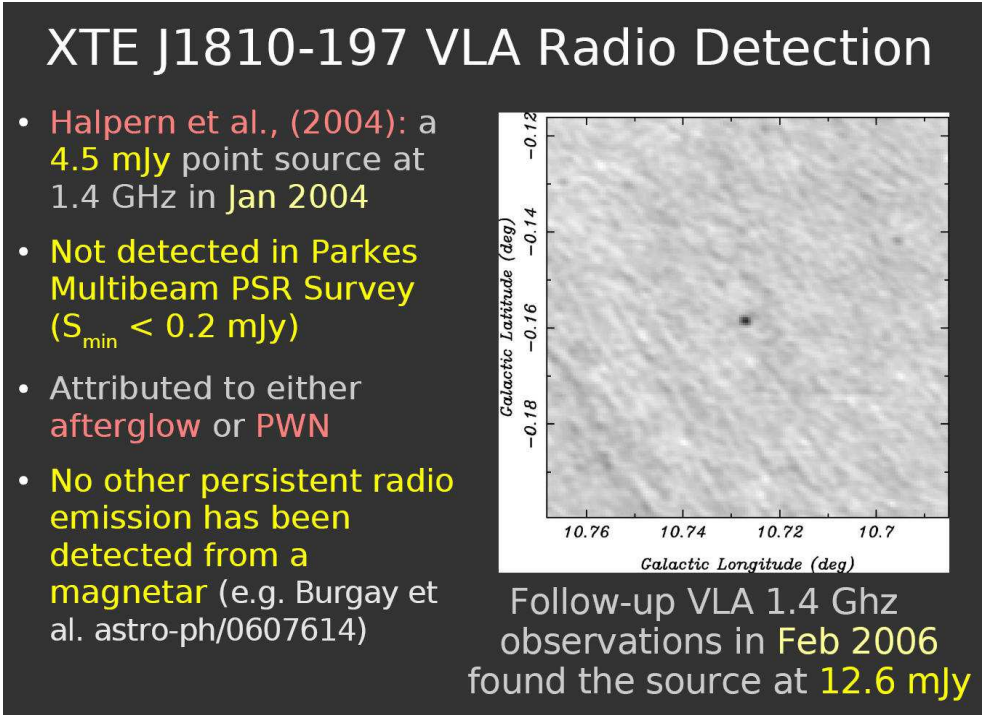,width=13cm,clip=}}
  \caption[]{\small Chart on the radio detection of XTE J1810-197 as
  presented in the talk of S.M.~Ransom.}
  \end{figure*}

\subsection{Future radio observatories for pulsar studies}\smallskip
{\bf M.~Kramer:} Over the next decade, radio astronomers will have new, exciting instruments
available to answer fundamental questions in physics and astrophysics. Without doubts, new
discoveries will be made, revealing new objects and phenomena. We can expect pulsar astronomers
to receive their fair share. In many respects, the field of pulsar astrophysics will change, as
the science will not simply be a continuation of what has been done so successfully over the
past 40 years. Instead, the huge number of pulsars to be discovered and studied with an
unprecedented sensitivity will provide a step forward into exciting times. Instruments like
the Low Frequency Array (LOFAR), the Square Kilometer Array (SKA) and their powerful pathfinder
telescopes will enable a complete census of Galactic pulsars, ultimate tests of gravitational
physics, unique studies of the emission process and much more. This talk demonstrated the
potential of the future instruments by presenting highlights of the science to be conducted.

  \begin{figure*} 
  \centerline{\psfig{figure=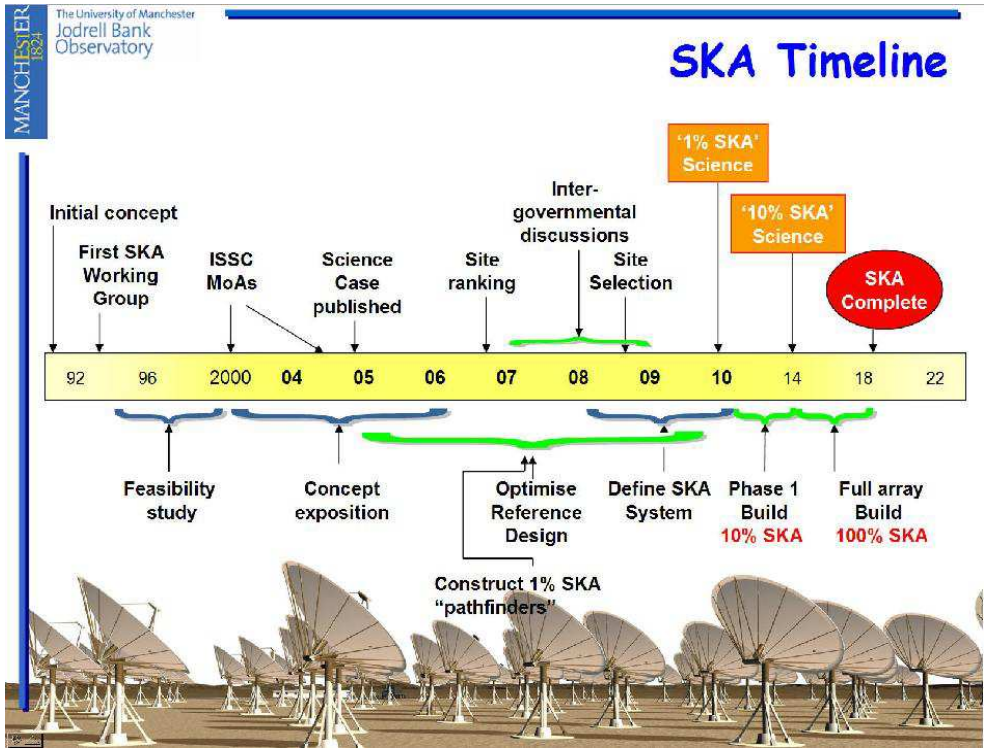,width=13cm,clip=}}
  \caption[]{\small Chart on the time line of the SKA as presented by M.~Kramer.}
  \end{figure*}

\subsection{Future optical and X-ray observatories for pulsar studies}\smallskip
{\bf W.~Becker:} Optical and X-ray astronomy has made great progress in the past several years
thanks to telescopes with larger effective areas and greatly improved spatial, temporal and spectral
resolutions. The next generation instruments like XEUS, Constellation-X, Simbol-X, eROSITA, the
James Webb Space Telescope and the ESOs Extremely Large optical Telescope are supposed to bring
again a major  improvement in sensitivity. The purpose of this talk was to summarize the future
plans for X-ray and optical telescopes with the emphasis of their application for pulsar and
neutron star astronomy.

  \begin{figure*} 
  \centerline{\psfig{figure=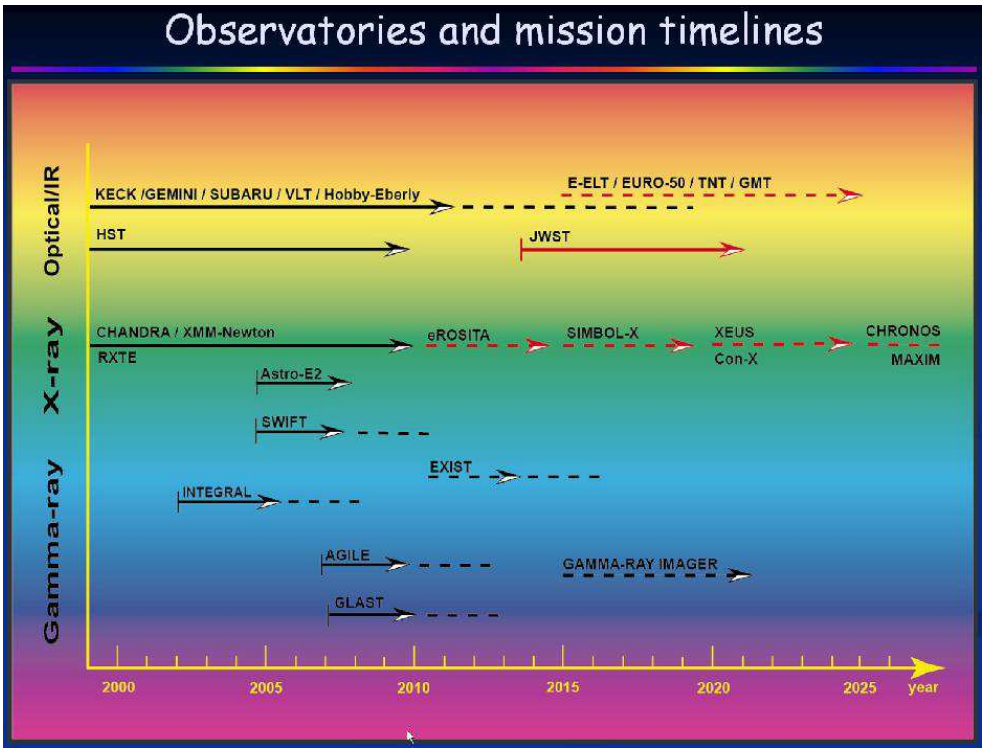,width=13cm,clip=}}
  \caption[]{\small Chart on the time line of future high energy missions
  as presented by W.~Becker.}
  \end{figure*}

\subsection{Future Gamma-ray and TeV observatories for pulsar searches}\smallskip
{\bf D.A. Smith:} GeV measurements of pulsar lightcurves versus energy provide information
on beam geometry, and spectral cut-offs give insights into the particle acceleration region(s)
around the neutron star. The two together can help build a better picture of neutron star
populations in the galaxy. We are on the verge of instrumental break-through that promise
to increase the sample of measured objects by ~10. This talk described instruments able 
to detect multi-GeV pulsations for the next few years. Developments of ground-based atmospheric 
Cherenkov detectors were covered, AGILEs prospects were reviewed, and then the talk focused 
on the LAT (Large Area Telescope) on GLAST, including sensitivity estimations. Accurate radio 
ephemerides can greatly enhance the gamma-ray pulsar science. Efforts to build a large
ephemerides database were described.

\section{Poster Contributions}\bigskip

\subsection{Effects of core magnetic fields in evolution of binary neutron stars}\smallskip
{\bf T.~Mirtorabi, A.J.~Khasraghi, S.~Abdolrahimi:} The standard scenario for evolution in 
a close binary system in which the neutron star pass through four evolutionary phases 
(isolated star, propeller, wind accretion and Roche lobe accretion) was employed to 
calculate transport of orbital angular momentum to the neutron star or loss of angular 
momentum from the whole system.
The evolution of  core magnetic field of  neutron stars  in  close binary systems with a low
mass main sequence companion was explored. Assuming the core as a type II superconductor so
the magnetic flux can be transported as quantized fluxoids, calculation have been performed
to determine magnetic filed decay  and its interaction with the mater accreted from the
companion. The evolution of semi major axis of the binary in a time scale of $10^9$ years 
comparable with the main sequence life time of the low mass companion was also investigated.

\subsection{External electromagnetic fields of slowly rotating relativistic magnetized NUT stars}\smallskip
{\bf B.J.~Ahmedov, A.V.~Khugaev:} Analytic general relativistic expressions for the electromagnetic
fields external to a slowly-rotating magnetized NUT star with non-vanishing gravitomagnetic charge
have been presented. Solutions for the electric and magnetic fields have been found after separating
the Maxwell equations in the external background spacetime of a slowly rotating NUT star into angular
and radial parts in the lowest order approximation in specific angular momentum and NUT parameter.
The relativistic star was considered isolated and in vacuum, with different models for stellar magnetic
field: i) mono-polar magnetic field and ii) dipolar magnetic field aligned with the axis of rotation.

It was shown that the general relativistic corrections due to the dragging of reference frames and 
gravitomagnetic charge were not present in the form of the magnetic fields but emerge only in the form 
of the electric fields. In particular, it was argued that the frame-dragging and gravitomagnetic
charge provide an additional induced electric field which is analogous to the one introduced by the 
rotation of the star in the flat spacetime limit.

\subsection{PSR J0538+2817 as the remnant of the first supernova explosion in a massive binary}\smallskip
{\bf V.V.~Gvaramadze:} It is generally accepted that the radio pulsar PSR J0538+2817 is associated
with the supernova remnant (SNR) S147. The only problem for the association is the obvious discrepancy
(Kramer et al.~2003) between the kinematic age of the system of ~30 kyr (estimated from the angular
offset of the pulsar from the geometric center of the SNR and pulsar's proper motion) and the
characteristic age of the pulsar of  ~600 kyr. To reconcile these ages one can assume that the pulsar
was born with a spin period close to the present one (Kramer et al.~2003; Romani \& Ng 2003).

An alternative explanation of the age discrepancy based on the fact that J0538+2817 could be the 
stellar remnant of the first supernova explosion in a massive binary system and therefore could 
be as old as indicated by its characteristic age was proposed. The proposal implied that S147 is 
the diffuse remnant of the second supernova explosion (that disrupted the binary system) and that 
a much younger second neutron star (not necessarily manifesting itself as a radio pulsar) should 
be associated with S147. The existing observational data on the system PSR J0538+2817/SNR S147 were
used to suggest that the progenitor of the supernova that formed S147 was a Wolf-Rayet star (so that 
the supernova explosion occurred within a wind bubble surrounded by a massive shell) and to constrain 
the parameters of the binary system. The magnitude and direction of the kick velocity received by the 
young neutron star at birth was also restricted and it was found that the kick vector should not 
strongly deviate from the orbital plane of the binary system.

 \begin{figure*}
  \centerline{\psfig{figure=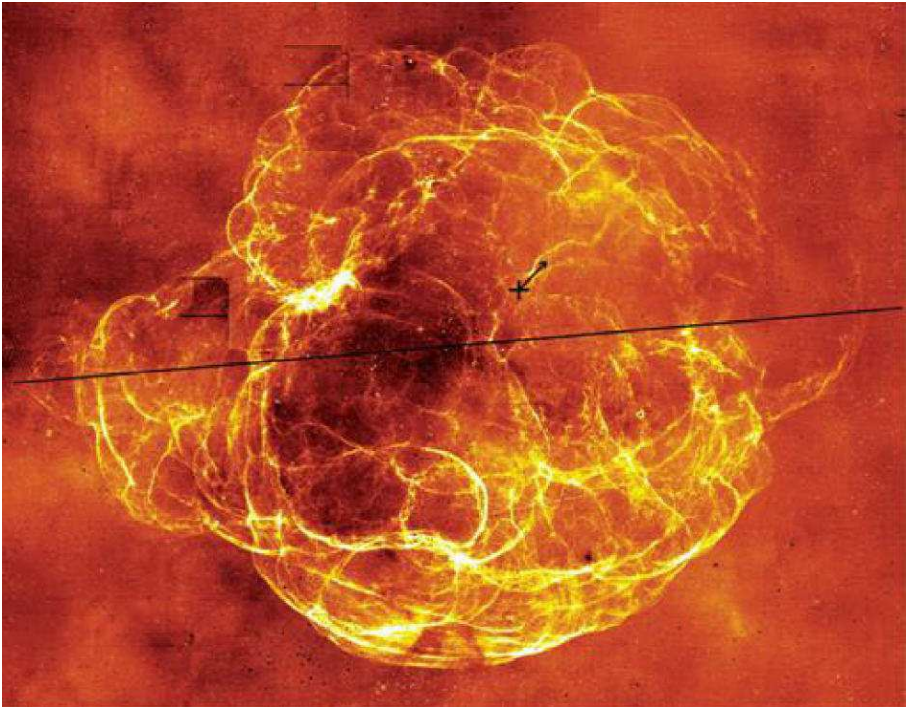,width=12cm,clip=}}
  \caption{H$_\alpha$ image of the SNR S147 (Drew et al.~2005; reproduced
  in the poster with the permission of the IPHAS collaboration). Position of PSR
  J0538+2817 is indicated by a cross. The arrow shows the direction of the
  pulsar proper motion vector (Kramer et al.~2003). The line drawn in the
  east-west direction shows the bilateral symmetry axis. North it up, east
  is west. Image from the poster 3.3 of V.V.~Gvaramadze.}
  \end{figure*}

\subsection{Ensemble pulsar time scale}\smallskip
{\bf A.E. Rodin:} The purpose of this work was to construct an algorithm of a new astronomical time 
scale based on the rotation of pulsars, which has comparable accuracy with the most precise 
terrestrial time scale TT.

This algorithm is based on the Wiener optimal filtering method and allows separating contributions
to the post-fit pulsar timing residuals of the atomic clock used in pulsar timing and spin variations
of the pulsar itself. The optimal filters were constructed with use of the cross and auto covariance
functions (in time domain) or auto and cross power spectra (in frequency domain) of the post-fit
timing residuals of pulsars participating in construction of the ensemble time scale.

The algorithm was applied to the timing data of millisecond pulsars PSR B1855+09 and B1937+21
(Kaspi et al.~1994) and allowed to obtain the corrections of UTC scale relative to ensemble
pulsar time scale PTens. Comparison of the differences UTC  PTens   and UTC  TT displays significant
correlation between them at level 0.79. Subsequent analysis of TT and  PTens shows that TT coincides with
PTens within 0.40+/-0.17 mcs and has a fractional stability $10^{-15}$ at the 7 years time interval.

Relatively close angular distance (15.5 degrees) on the sky between these pulsars gives grounds to
expect that there is a correlated signal in the post-fit timing data caused by the stochastic
gravitational wave background (GWB). A new limit of the fractional energy density of GWB based on the
difference TT and  PTens was established to be $\Omega_g h^2 \sim 2\times 10^{-10}$. This new value is
by one order lower than the previously published one owing to application of the new algorithm that
separates the proper pulsars spin and local atomic standard variations.

\subsection{Comparison of giant radio pulses in young pulsars and millisecond pulsars}\smallskip
{\bf A.~Slowikowska, A.~Jessner, G.~Kanbach, B.~Klein:} Pulse-to-pulse intensity variations are a common 
property of pulsar radio emission. For some of the objects single pulses are often 10-times stronger than 
their average pulse. The most dramatic events are the so called giant radio pulses (GRPs). They can be 
1000-times stronger than the regular single pulses from the pulsar. Giant pulses are a rare phenomenon, 
occurring in very few pulsars which split into two groups. The first group contains very young and energetic 
pulsars like the Crab pulsar, and its twin in the LMC (PSR B0540-69), while the second group is represented 
by old, recycled millisecond pulsars like PSR B1937+21, PSR B1821-24, PSR B1957+20, and PSR J0218+4232 - the
only millisecond pulsar detected in gamma-rays. The characteristics of GRP's for these two pulsar groups was
discussed. In particular, the poster focused on the flux distributions of GRPs which were compared. Moreover, 
the latest findings of new features in the Crab GRPs were presented. Analysis of Effelsberg data taken at 
8.35~GHz have shown that GRPs do occur in all phases of its ordinary radio emission, including the phases 
of the two high frequency components (HFCs) visible only between 5 and 9~GHz. This suggests that a similar 
emission mechanism may be responsible for the main pulse, the inter pulse and the HFCs. Finally, the similarities 
and differences between both groups of pulsars in the context of timing, spectral and polarization properties 
of these pulsars were discussed. It was also attempted to answer the question why pulsars belonging to so 
different classes do show the same giant radio emission phenomena.

\subsection{Integral IBIS and JEM-X observations of PSR B0540-69}\smallskip
{\bf A.~Slowikowska, G.~Kanbach, J.~Borkowski, W.~Becker:} The high-energy pulsar PSR B0540-69 in the Large 
Magellanic Cloud (d $\sim$ 49.4 kpc), embedded in a synchrotron plerion in the center of SNR 0540-69.3 is often 
referred to as an extragalactic 'twin' of the Crab pulsar. Its pulsed emission has been detected up to about 
48 keV so far. The results from the search for PSR B0540-69 up to 300 keV in the 1 Ms INTEGRAL data was presented. 
INTEGRAL was pointed to the LMC during 6 revolutions in January 2003, and 3 revolutions in January 2004. The 
events used for timing and spectral analysis of the source come from the IBIS/ISGRI and JEM-X detectors. 
Moreover, the details of data analysis technique used for this weak source were presented.

\subsection{Plasma modes along open field lines of neutron star endowed with gravitomagnetic NUT charge}
{\bf B.J.~Ahmedov, V.G.~Kagmanova: }\smallskip
Electrostatic plasma modes along the open field lines of a rotating neutron star endowed with
gravitomagnetic charge or NUT parameter have been considered. Goldreich-Julian charge density in
general relativity was analyzed for the neutron star with non-zero NUT parameter. It was found that 
the charge density is maximal at the polar cap and remains almost the same in a certain extended region
of the pole. For a steady state Goldreich-Julian charge density it was found that the usual plasma 
oscillation along the field lines; plasma frequency resembles the gravitational redshift close to 
the Schwarzschild radius. The results in studying the nonlinear plasma mode along the field lines were
presented. The equation contained a term that described the growing plasma modes near Schwarzschild 
radius in a black hole environment. The term vanished with the distance far away from the gravitating 
object. For initially zero potential and field on the surface of a neutron star, Goldreich-Julian charge 
density was found to create the plasma mode which was enhanced and propagated almost without damping 
along the open field lines of magnetized NUT star.

\subsection{The drift model of magnetars}\smallskip
{\bf I.F.~Malov, G.Z.~Machabeli:} It was shown that the drift waves near the light cylinder can cause the 
modulation of emission with periods of order several seconds. These periods explain the intervals between 
successive pulses observed in "magnetars" and radio pulsars with long periods. The model under consideration  
gave the possibility to calculate real rotation periods of host neutron stars. They are less than 1 sec for 
the investigated objects. The magnetic fields at the surface of the neutron star are of order $10^{11}$ - $10^{13}$ G 
and equal to the fields  usual for known radio pulsars.

\subsection{Sub-stellar companions around neutron stars}\smallskip
{\bf B.~Posselt, R.~Neuh\"auser, F.~Haberl:} Planets or sub-stellar companions around neutron stars
can give valuable insights into a neutron star's formation history considering for example birth
kicks or fallback disks. They may also help to derive neutron star masses which would be very
welcome especially if the radius can be derived by other means as for the radio-quiet X-ray
thermal neutron stars. Currently there are two planetary systems around millisecond pulsars known.
They have been found by the pulse timing technique which is most sensitive to old  millisecond
pulsars. Some of the formation theories can already be ruled out for these systems. However,
statistics are very poor and other search techniques are needed to cover also young, even
radio-quiet neutron stars.

The first results of direct imaging search for sub-stellar companions around the closest and 
youngest neutron stars started three years ago with ESO's VLT were presented. Among the objects 
was the famous RX J1856.5-3754 for which a sub-stellar object could help to constrain the equation 
of state as the radius was already previously derived by its X-ray thermal emission.

  \begin{figure*}[t!]
  \centerline{\psfig{figure=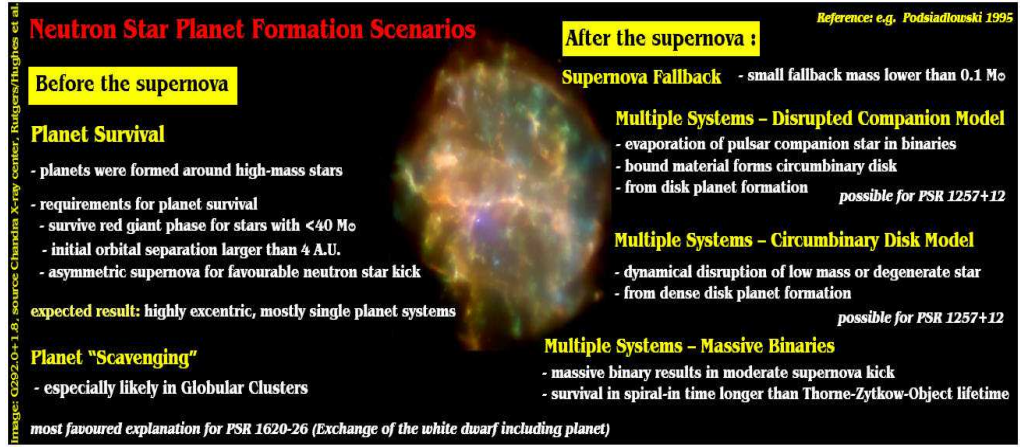,width=13.5cm,height=6cm,clip=}}
  \caption{Image from the poster 3.9 of Posselt, Neuh\"auser and Haberl.}
  \end{figure*}

\subsection{Detection of the individual pulses of the pulsars B0809+74, B0834+06, B0950+08,
B0943+10, B1133+16 at decameter wave range}\smallskip {\bf O.M.~Ulyanov et al.:} Radio emission of 
single pulses for five pulsars was found at frequencies $18-30$ MHz.  It was reported that the radio 
emission is caused by the strong subpulses that have peak intensity of more than 20 times larger 
than the peak intensity of average profiles. It was found that the intensity of single pulses has 
a strong variation in frequency and time.  The probability of detection of the anomalous intense 
pulses thus does not exceed several percents at Decameter wave range. Usually such pulses are 
detected in short series (not more than 10 pulses). Typical band values of detection for the 
pulses with anomalous intensities lies in the range from 0.2 to 0.5 octaves.

\subsection{An analytical description of low-energy secondary plasma particle distribution in pulsar magnetospheres}\smallskip
{\bf V.M.~Kontorovich, A.B.~Flanchik:} A simple analytical approximation of the form of low-energy 
cut-off of the secondary particle distribution was proposed. This approximation is acceptable for 
describing the known cascade numerical simulation. The distribution form and maximum position as
function of pulsar parameters have been found theoretically by considering the curvature radiation 
process and the electron-positron pairs production. An influence of synchrotron radiation on the 
form and maximum of the low-energy distribution was considered.

\subsection{Timing irregularities and the neutron star stability}\smallskip
{\bf J.O.~Urama: } Observations show that the different manifestations of neutron stars exhibit 
measurable departures from the predicted slow down. This has been largely attributed to rotational 
irregularities, timing noise, glitches, and precession. The analysis of the regular spin-down and 
jump parameters for a combined set of the pulsars (radio, optical, x-ray and gamma-ray) and magnetars 
was presented. It was also attempted to quantify the stability of neutron stars in general.

\subsection{Force-free pulsar magnetosphere}\smallskip
{\bf A.N.~Timokhin: } The properties of a force-free pulsar magnetosphere and address the role of 
electron-positron cascades in determining a particular configuration among other possible force-free 
magnetospheric configurations were discussed. Results of high resolution numerical simulations of the 
force-free magnetosphere of aligned rotator and analyze in details properties of an aligned pulsar 
were reported. It was argued that the closed field line zone should grow with time slower than the 
light cylinder; this yield the pulsar braking index less than 3. However, models of aligned rotator 
magnetosphere with widely accepted configuration of magnetic field, when the last closed field line 
lies in equatorial plane at large distances from pulsar, have serious difficulties. The solutions 
of this problem were discussed and it was argued that in any case, also for inclined pulsar, 
energy losses should evolve with time differently than it is predicted by the magneto-dipolar formula 
and the pulsar braking index should be different from the "canonical" value equal to 3.

\subsection{Crab pulsar optical photometry and spectroscopy with microsecond temporal resolution}\smallskip
{\bf G.~Beskin et al.: } The results of fast photometry and spectroscopy of the Crab pulsar
with microsecond temporal resolution were presented. The observations have been performed on
William Herschel 4-m and BTA 6-m telescopes using APD avalanche photon counter and PSD panoramic
photon imager.

The stability of the optical pulse was analyzed and the search for the variations of the pulse
shape along with its arrival time stability was performed. Upper limits on the possible short
time scale free precession of the pulsar and the stochastic variable optical emission component
were placed. The results of the low resolution ($\sim 300$ angstrom) phase-resolved spectroscopy of
the pulsar emission were discussed, first of all  the distinction of the spectra of pulses and
off-pulse phase intervals.

\subsection{Discovery of a large time scale cyclic evolution of radio pulsars rotational frequency}\smallskip
{\bf G.~Beskin, A.~Biryukov, S.~Karpov: } The recent massive measurements of  pulsar frequency second 
derivatives have shown that they are $100-1000$ times larger than expected for standard pulsar 
slowdown. Moreover, the second derivatives as well as braking indices were even negative for 
about half of the pulsars. It was explained that these paradoxical results from statistical analysis
of the rotational parameters (frequency, its first and second derivatives) 
of the subset of 295 pulsars taken mostly from the ATNF database. The strong correlation of second 
and first frequency derivatives either for positive (correlation coefficient r $\sim$ 0.9) and 
negative (r $\sim$ 0.85) values of second derivative, and of the frequency and its first 
derivative (r$\sim$ 0.7) were found. These dependencies were interpreted as evolutionary  ones  
due to the first frequency derivative being nearly proportional to the characteristic age. The 
derived statistical relations as well as "anomalous" values of the second frequency derivative 
were well explained in the framework of the simple model of cyclic evolution of the rotational 
frequency of the pulsars. It combined the secular change of the rotational parameters according 
to the power law with braking index n$\sim$ 5 and harmonic oscillations of 100--1000 years period 
with an amplitude from 10-3 Hz for young pulsars to 10-10 Hz for older ones. It was found that 
the physical nature of these cyclic variations of the rotational frequency may be similar to 
the well-known red timing noise, however, with much larger characteristic time scale.

  \begin{figure*}[t!]
  \centerline{\psfig{figure=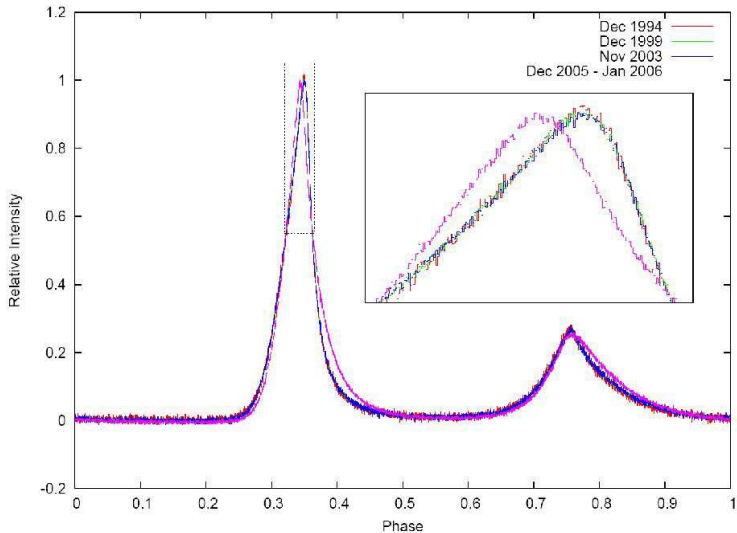,width=10cm, clip=}}
  \caption{Phase aligned light curves of the Crab pulsar from optical observations at 
  different epochs all scaled to the same pulse height. In the observation at the 6-meter 
  SAO telescope in January 2007 the Crab pulsar light curve was similar to
  the one observed in 1994, 1999 and 2003. This means that the change of the light curve 
  observed in December 2005 / January 2006 was either a stochastic (rare) event or a 
  not understood instrumental effect. Image from the poster 3.14 of G.~Beskin et al.}
  \end{figure*}

\subsection{Abnormal phases in nuclear matter in supernova core collapse model}\smallskip
{\bf D.J.~Bora, H.L.~Duorah, K.~Duorah: } The role of abnormal phases in nuclear matter on the basis
of the well-known Lee-Wick theory was studied for the determination of the shock strength at the
core bouncing of type II supernova. Relativistic equation for collapse beyond nuclear density with
a still stronger magnetic field was developed to study the effect of appearance of the abnormal
phases. This lead to the softening of the equation of state giving rise to strong shock. The
magnetar thus formed can be a store-house of many high energy events including rapid gamma-ray
bursts. A strong magnetic field beyond some critical value sometimes can exponentially accelerate
certain reactions  producing rapid bursts of energy. Some seismic events on the solidified
crusts of magnetars can also be alluded for the short gamma-ray burst discovered in December 2004
at the site of SGR 1806-20.

\subsection{Instant radio spectra of giant pulses from the Crab pulsar over decimeter to decameter
wave bands}\smallskip {\bf M.V.~Popov et al.: } The results of simultaneous multi-frequency observations 
of giant radio pulses (GPs) from the Crab pulsar PSR B0531+21 at frequencies of 23, 111 and 600 MHz were
presented. For the first time GPs were detected at such low frequency as 23 MHz. Among 45 GPs detected 
in the overall observations time with 600 MHz, 12 GPs were identified as simultaneous ones at 600 and 
23 MHz. At 111 MHz among 128 GPs detected in the overall observations time with 600 MHz, 21 GPs were 
identified as simultaneous ones at 600 and 111 MHz. Spectral indices for the power-law frequency 
dependence of GPs energy were enclosed between -3.1 and -1.6. Mean spectral index equals to $-2.7\pm 0.1$ 
and was the same for both frequency combinations 600-111 MHz and 600-23 MHz.

A big scatter in values of the individual spectral indexes and a large number of unidentified 
giant pulses indicated that a real form of spectra of individual giant pulses did not follow 
a simple power law. The shape of giant pulses at all three frequencies was governed by the 
scattering of radio waves on the inhomogeneities of the interstellar plasma. The pulse scatter 
broadening and their frequency dependence was measured as 20(n/100)-3.50.1 ms, where frequency 
n is in MHz.

\subsection{Pulsar nulling quantitative analysis}\smallskip
{\bf J.H.~Seiradakis, K.~Lazaridis: } Using long sequences of single pulses the nulling behavior 
of several pulsars was analyzed. Each pulsar has been characterized by a "nulling parameter", which 
represents the average length of consecutive null pulses and a "nulling max" parameter, which
represents the maximum length of consecutive null pulses. These two parameters were compared to 
other pulsar parameters. Some interesting correlations were derived.

\subsection{Eclipse study of the double pulsar}\smallskip
{\bf R.P.~Breton et al.:} The double pulsar system PSR J0737-3039 offers an unprecedented opportunity 
for studying General Relativity and neutron-star magnetospheres. This system has a favorable orbital 
inclination such that the millisecond pulsar, "A", is eclipsed when its slower companion, "B", passes 
in front. High time resolution light curves of the eclipses reveal periodic modulations of the radio 
flux corresponding to the fundamental and the first harmonic of pulsar "B" spin frequency. Eclipse 
modeling is highly sensitive to the geometrical configuration of the system and thus provides a 
unique probe for parameters like the inclination angle of pulsar "B" spin axis as well as their time 
evolution due to relativistic effects. Detailed fitting of the pulsar "A" eclipse light curves to 
a model that includes, for pulsar "B", a simple dipolar magnetic field was presented. It was found 
that the eclipses can be reproduced very well, and one obtains precise measurements of pulsar "B's" 
orientation in space. Results on a search for secular changes caused by geodetic precession of pulsar 
"B's" spin axis were reported.

\subsection{Observations of southern pulsars at high radio frequencies}\smallskip
{\bf A.~Karastergiou, S.~Johnston: } A number of pulsars at 1.4, 3.1 and 8.4 GHz in full polarization 
at the Parkes radio-telescope was observed. The main objective was to study the frequency evolution of 
polarization by means of new high quality polarization data at high frequencies. Average polarization 
profiles with high time resolution to update already existing previous observations were also obtained. 
Detailed description of a total of 97 polarization profiles at the 3 aforementioned frequencies was 
provided. This relatively large sample provided the opportunity to study effects related to the linear 
and circular polarization as well as the polarization position angle. An evidence was found that: 1) a 
simple model where two orthogonal polarization modes with competing spectral indices can account for 
many observational properties of the linear polarization and total power, 2) the position angle 
dependence on frequency depends on the different relative strength of profile components at different 
frequencies, 3) young, energetic pulsars remain highly polarized at high frequencies, and 4) highly 
polarized components may originate from higher up in the pulsar magnetosphere then unpolarized 
components.  A summary of these results was presented.

\subsection{Pulsar braking indices}\smallskip
{\bf A.~Baykal, A.~Alpar: } Almost all pulsars with anomalous positive second derivative of angular 
acceleration measurements (corresponding to anomalous braking indices in the range $5<n<100$), 
including all the pulsars with observed large glitches as well as post glitch or inter-glitch
second  measurement obey the scaling between glitch parameters originally noted in the Vela 
pulsar. Negative second derivative values can be understood in terms of glitches that were 
missed or remained unresolved. The glitch rates and a priori probabilities of positive and 
negative braking indices according to the model developed for the Vela pulsar were discussed. 
This behavior supports the universal occurrence of a nonlinear dynamical coupling between 
the neutron star crust and an interior superfluid component.

\subsection{Electrodynamics of pulsar's electrosphere}\smallskip
{\bf J.A.~Petri: } A self-consistent model of the magnetosphere of inactive, charged, aligned 
rotator pulsars with help of a semi-analytical and numerical algorithm, was presented. In this
model the only free parameter was the total charge of the system.  This "electro-sphere" is 
stable to vacuum breakdown by electron-positron pair production.  However, it appears to be 
unstable to the so-called ``diocotron'' instability which is an electrostatic instability.  
Eigenspectra and eigenfunctions for different disc models, which differ by the total charge 
of the disc-star system were presented. The evolution of this instability on a long time-scale 
was studied in a fully non-linear description by means of numerical simulations. For multi-mode
excitation, the average macroscopic response of the system could be described by a quasi-linear 
model.  It was found that in the presence of an external source feeding the disk with positive 
charges, representing the effect of pair creation activity in the gaps, the diocotron instability 
may give rise to an efficient diffusion of charged particles across the magnetic field lines.

 \begin{figure*}[h!]
  \centerline{\psfig{figure=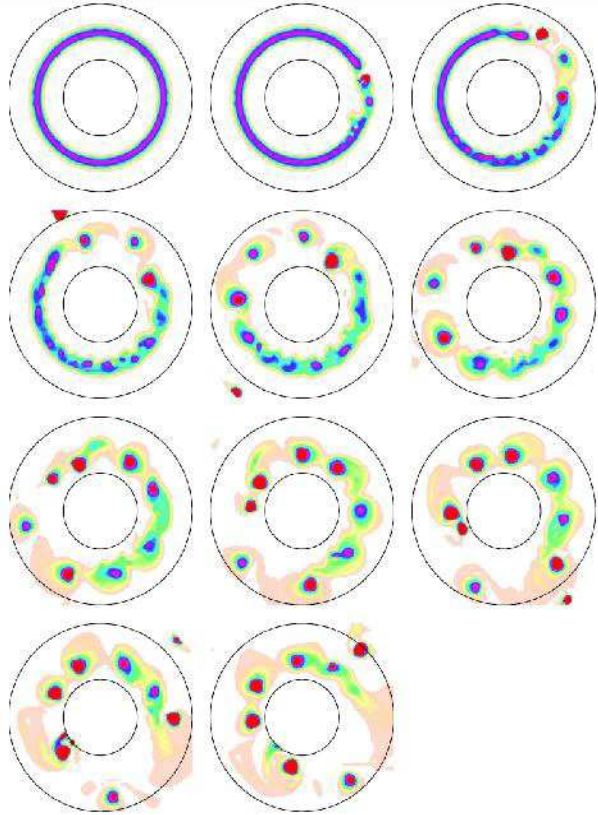,width=8cm, clip=}}
  \caption{Non-linear evolution of the diocotron instability in an
  infinitely thin disk. Image from the poster 3.22 of J.A.~Petri.}
  \end{figure*}

\subsection{About one hypothesis on the origin of anomalous X-ray pulsars and soft
Gamma-ray repeaters}\smallskip {\bf F.~Kasumov, A.~Allakhverdiev, A.~Asvarov: } The possibility 
of realization of the scenario, according to which anomalous X-ray pulsars (AXPs) and soft 
Gamma-ray repeaters (SGRs) originate from the radio pulsars with the very close initial 
parameters (period, magnetic field etc.), subjected to considerable and prolonged 
glitches, was analyzed. This scenario provides both an increase in the period of ordinary
pulsars and the attainment of magnetic field strengths typical of these objects 
(B $\sim 10^{13} - 10^{14}$ G), a new class of neutron stars, called magnetars, at an 
insignificant initial magnetic field value B $\sim 3\times 10^{10} -10^{11}$ G. 
With this aim, the criteria to be satisfied by a potential progenitor of AXPs and SGRs
were determined and analyzed. So, taking into account the combined action of all factors 
(magnetic field, distance, birth place, satisfying to our criteria etc.) we restricted 
our analysis to ~100 pulsars with $B > 5\times 10^{12}$ G and $P > 0.5$s. The observed 
characteristics of such pulsars, their association with supernova remnants, and their 
evolution in the $P-\dot P$ diagram with allowance made for the actual age of the possible 
AXP and SGR progenitor were shown to be in conflict with the suggested scenario and can be
better described in the framework of the standard magneto-dipole model of pulsar evolution.

\subsection{Combined models of evolution and real ages of pulsars}\smallskip
{\bf A.~Allakhverdiev, F.~Kasumov, S.~Tagieva: } The test for checking the applicability 
of fall-back or propeller models for the pulsar evolution was proposed. This test was 
based on the comparison of the pulsars ages predicted by these models with the real 
kinematical ages of pulsars. With this goal two groups of pulsars, namely relatively 
young and old pulsars with the distances from the galactic plane $z < 100$ pc and 
$z < 300$ pc, respectively, were selected. At the same time, the irregular character 
of deviation of the pulsars birthplaces from the geometrical plane of the Galaxy has 
been taken into account. The distribution of these groups of pulsars in Period-Period 
derivative diagram was compared with the theoretical tracks of the evolution of the 
pulsars predicted by fall-back and combined dipole $+$ propeller models at various 
values of the initial parameters of radio pulsars (magnetic field and accretion rate 
of matter). As it is well known, characteristic feature of combined model (unlike to 
the pure magneto-dipole) is the increase of period derivative up to some critical 
value with increase of the period of pulsars, i.e.~the age of pulsars. It was shown 
that the distribution of selected pulsars in the mentioned diagram contradicts to 
this model and can be easily explained by standard model of pulsar evolution.

\subsection{X-ray emission from hot polar cap in pulsars with drifting subpulses}\smallskip
{\bf J.~Gil, G.~Melikidze: } Within the framework of the partially screened inner 
acceleration region the relationship between the X-ray luminosity and the circulational 
periodicity of drifting subpulses was derived. This relationship was quite well satisfied 
in pulsars for which an appropriate radio and X-ray measurements exist. A special case of 
PSR B0943+10 was presented and discussed. The problem of formation of a partially screened 
inner acceleration region for all pulsars with drifting subpulses was also considered. 
It was argued that an efficient inner acceleration region just above the polar cap 
can be formed in a very strong and curved non-dipolar surface magnetic field.

\subsection{Pulsed radio emission from two XDINS}\smallskip
{\bf V.M.~Malofeev, O.I.~Malov, D.A.~Teplykh: } Investigations of two X-ray dim isolated 
neutron stars: J1308.6 +212708 and J2143.03+065419 were reported. The observations were carried 
out on two sensitive transit radio telescopes in Pushchino at a few frequencies in the range 
$111-42$ MHz. Mean pulse profiles, the flux density and the dispersion measures were presented. 
The measures of periods and their derivatives were reported, as well as the estimation of 
distances and integral radio luminosities. The comparison with X-ray observations was made.

\subsection{Magnetospheric eclipses in the double pulsar system J0737-3039}\smallskip
{\bf R.R.~Rafikov, P.~Goldreich: } The recently discovered double pulsar system, PSR J0737-3039, 
consisting of a millisecond and a normal pulsars in a 2.4 hour orbit, provides us with 
unprecedented tests of general relativity and magnetospheric effects.  One of the most interesting 
phenomena observed in this system is the eclipse of the millisecond pulsar in the radio at its 
conjunction with the normal pulsar. A theory which explains this observation as a result of 
synchrotron absorption of the millisecond pulsar radio beam in the magnetosphere of the normal 
pulsar was presented. Absorption was induced in a sense that the intense radio beam of the 
millisecond pulsar itself strongly modifies the properties of the plasma in the closed part 
of the normal pulsar magnetosphere: absorption of high-brightness temperature radio emission 
heats up particles already present there and also allows additional pair plasma to be trapped 
in this region by magnetic bottling effect. This theory self-consistently predicts the size 
of the eclipsing region which agrees very well with the observed duration of eclipse. 
Recent observations of the variability of transmission during the eclipse modulated at the
rotation period of the normal pulsar have been interpreted as resulting from the absorption 
by the rigidly rotating dipolar-shaped magnetosphere which is in perfect agreement with the 
presented theory.

\subsection{Electromagnetic fields of magnetized neutron stars in braneworld}\smallskip
{\bf B.J.~Ahmedov, F.J.~Fattoyev: } The dipolar magnetic field configuration in dependence 
on brane tension and present solutions of Maxwell equations in the internal and external 
background spacetime of a magnetized spherical neutron star in a Randall-Sundrum II type 
braneworld was studied. The star was modeled as a sphere consisting of perfect highly 
magnetized fluid with infinite conductivity and frozen-in dipolar magnetic field. With 
respect to solutions for magnetic fields found in the Schwarzschild spacetime brane 
tension introduces enhancing corrections both to the interior and the exterior magnetic 
field. These corrections could be relevant for the magnetic fields of magnetized compact 
objects as pulsars and magnetars and may provide the observational evidence for the brane 
tension through the modification of formula for magneto-dipolar emission which gives 
amplification of electromagnetic energy loss up to few orders depending on the value 
of the brane tension.

\subsection{On dependence of some parameters of radio pulsars radiation on their age}\smallskip
{\bf V.H.~Malumian, A.N.~Harutyunyan: } The relationship between parameters of the 
radiation from pulsars and the dependence of the rates of the radiation periods of 
these objects on their characteristic ages was studied. The following results were 
obtained:

(a.) The rate of change in the radiation periods (derivatives of periods, dP/dt) of 
pulsars depends on their characteristic age. These changes proceed more slowly with 
age. The rate of change of the radiation period of pulsars can in some way serve 
as an indicator of their age.

(b.) The relationship between the rate dP/dt of change of the period and the period P has been
demonstrated. For young pulsars this relationship is weak. In the course of evolution with age,
the relationship between the derivative of the period and the period becomes closer. Whereas for
young ($T < 10^6$ years) pulsars the correlation coefficient for the log dP/dt -- log P plot is
only  0.49 ($p<0.0001$), for old ($T \ge 10^8$ years) pulsars the correlation
coefficient approaches unity ($p<0.0001$). Since, as shown above, the rate of change of the radiation
period decreases with age in pulsars, one can say that the lower the rate of change of the
radiation period of a pulsar is, the closer is the relationship between the derivative of its
period and its period.

The data in the catalog of Taylor et al., which contains 706 objects, were examined 
separately. After eliminating the members of binary and multiple systems, as well 
as the members of the Magellanic Clouds, slightly more than 500 objects remain.

\subsection{The Nancy pulsar instrumentation: The BON coherent dedispersor}\smallskip
{\bf I.~Cognard, G.~Theureau: } A summary of the Nancay pulsar instrumentation and 
the on going observational pulsar timing programs was presented.
      The BON coherent dedispersor is able to handle 128MHz of bandwidth. It is made of a spectrometer,
plus four data servers to spread data out to a 70-node cluster of PCs (with Linux Operating System).
De-dispersion is done by applying a special filter in the complex Fourier domain. This backend has been
designed in close collaboration with the UC Berkeley. It benefits from the many qualities of the large Nancy
radio telescope (NRT, equivalent to a 94 m circular dish), which receivers were upgraded in 2000: a
factor  of 2.2 sensitivity improvement was obtained at 1.4MHz, with an efficiency of 1.4K/Jy for a
system temperature of 35K; a better frequency coverage was also achieved (from 1.1 to 3.5GHz).
     The first two years of BON data acquisition demonstrates that the timing data quality is comparable
with the Arecibo and Green Bank results. As an example, a Time Of Arrival (TOA) measurement accuracy better
than 200ns (170-180ns) is obtained in only 30 seconds of integration on the millisecond pulsar PSR B1937+21.
With this up to date instrumentation, two main observational programs in pulsar timing with the
Nancy antenna are operated: 1) the radio follow-up of X- and gamma-ray pulsars for the building of a complete
multi-wavelength sample and 2) the monitoring of both a millisecond pulsar timing array and a targeted list
of  binary or unstable pulsars for gravitational wave detection. Joining both list of targets, a total
sample of 150 pulsars is then monitored regularly with a dense sampling in time.

\subsection{Relation of pulsars to the remnants of supernova bursts}\smallskip
{\bf V.H.~Malumian, A.N.~Harutyunyan: } Based on a large volume of statistical data it was shown that
the spatial distributions of radio pulsars in the galaxy with characteristic ages $T < 10^6$ years and
$T > 10^6$ years differ significantly. The overwhelming majority of the pulsars with $T < 10^6$ years
lie within a narrow band of width 400 pc around the galactic plane. A large portion of the pulsars
with $T > 10^6$ years is concentrated outside this zone. In the case of younger pulsars, a
larger fraction of them lies within the confines of the above mentioned zone. It is also shown that
pulsars with $T < 10^6$ years and the remnants of supernova explosions have essentially the same spatial
distribution. These facts support the existence of a relationship between pulsars and
supernova remnants, as well as the acquisition of high spatial velocities by pulsars during their birth.

\subsection{The multi-photon electron-positron pair production in the magnetosphere of pulsars}\smallskip
{\bf A.K.~Avetissian: } In general, the single-photon reaction $\gamma \rightarrow  e^- + e^+$, as well 
as the inverse reaction of the electron-positron annihilation can proceed in a medium that must be 
a plasma-like. To provide a macroscopic refractive index $n(\omega)<1$ necessary for pair production 
$\gamma$-frequencies one needs plasma densities $\rho > 10^{33}\,\mbox{cm}^{-3}$. Such superdense matter 
exists in the core of the neutron stars/pulsars. At these densities the electron component of the 
superdense plasma is fully degenerated and taking also into account the Pauli principle the 
probabilities of these processes actually turn to zero. Hence, the possibility of multi-photon 
electron-positron pair production by strong electromagnetic radiation of soft frequencies in 
magnetosphere of pulsars is considered, which is possible at ordinary densities of plasma. Such
multi-photon process occurs via nonlinear channels at high intensities of electromagnetic radiation 
in wide region of frequencies from radio to UV and soft X-ray in pulsars magnetosphere.

Numerical simulations for various pulsars ($\Omega \sim 1 - 200$ s$^{-1}$) with the help of analytical 
distribution  functions of magnetospheres plasma with densities $\rho \sim 10^{20}- 10^{22}\mbox{cm}^{-3}$ 
and pair production probabilities have been made, and both energetic and angular distributions 
of produced electron-positron pairs were presented.

\subsection{Relativistic, electromagnetic waves in pulsar winds}\smallskip
{\bf O.~Skjaeraasen: } Extremely nonlinear, coherent electromagnetic waves in the context of relativistic, 
expanding plasma flows, where a confining external medium triggers the formation of a shock, were considered. 
Using a combination of analytical methods and Particle-In-Cell simulations, the mechanisms of wave generation 
and dissipation, as well as how the waves affect the particle distribution were discussed. For a large-amplitude 
wave of general polarization, any given set of wave parameters uniquely fixes the particle and energy flux 
associated with the flow. In cases where the wave properties can be constrained, this can be used to estimate 
the flow parameters.

The prime application of this work is to pulsar winds and pulsar wind termination shocks, where our model 
provides a viable alternative to magneto-hydrodynamic models. Using canonical parameters for the Crab, the 
mode couplings and transitions between the inner and outer parts of the wind were discussed. The simulation 
data were explored to shed new light on the microphysics of the wave as it reaches the shock.

\subsection{Coupled spin, mass, magnetic field, and orbital evolution of accreting neutron stars}\smallskip
{\bf M.~Mirtorabi, A.J.~Khasraghi, S.~Abdolrahimi: } The presented study was mainly addressed to the coupled 
spin, mass, magnetic field, orbital separation, and orbital period evolution of a neutron star entering 
a close binary system with a low mass main sequence companion, which loses mass in form of homogenous 
stellar wind. Flux expulsion of the magnetic field from the superfluid superconductive core of a neutron 
star, based on different equation of states was applied, and its subsequent decay in the crust, which also 
depends on conductivity of the crust, and hence on the temperature, T, and the neutron star age. The initial 
core and surface magnetic field were of the same order of magnitude. To derive the rate of expulsion of the
magnetic flux out of the core  various forces which act on the fluxoids in the interior of a neutron star 
were considered, including a force due to their pinning interaction with the moving neutron vortices, 
buoyancy force, curvature force, and viscous drag force due to magnetic scattering of electrons. Various 
effects accompanying mass exchange in binaries can influence the evolution of spin and magnetic field of 
the neutron star. The orbital separation of the binary clearly affects the estimated value of, and it 
itself evolves due to mass exchange between the components, mass loss from the system, and two other sinks 
of the orbital angular momentum namely magnetic braking and gravitational waves. The neutron star passes 
through four evolutionary phases (isolated pulsar- propeller- accretion from the wind of a companion- accretion 
resulting from Roche-lobe overflow). Models for a range of parameters, and initial orbital period, magnetic 
field and spin period were constructed. The impurity parameter, Q, was assumed to be constant during the 
whole evolution of the star and range from 1 to 0.001.  Final magnetic field, spin and orbital period were
presented in this paper. The suggested mechanism could
explain the lower magnetic field and faster spins of millisecond pulsars that have been recycled by 
accretion in close binaries.

\subsection{Investigating the magnetic field of the solar corona with pulsars}\smallskip
{\bf S.~Ord: } A novel experiment to examine both the magnetic field and electron content of the solar 
corona was proposed. It was intended to measure the Faraday rotation and dispersion evident in observations 
of background pulsar sources as they are occulted by the Sun. With a number of simultaneous lines of sight 
that cut different paths through the corona as the Sun rotates, strong constrains on the global topology 
of both the plasma and the magnetic field should be obtained. Although similar experiments have been performed 
using other background radio sources and space probes, this experiment differs in that many lines of sight 
can be examined simultaneously, and the magnetic field and plasma density can be measured
independently. The Parkes radio telescope is proposed to observe a number of pulsars as they are occulted 
by the Sun in December 2006. An outline of the experiment and a discussion of the expected results were 
presented.

\subsection{RRATs and PSR B1931+21}\smallskip
{\bf X.D.~Li: }  The recent discovery of rotating radio transients and the quasi-periodicity of pulsar 
activity in the radio pulsar PSR B1931+24 has challenged the conventional theory of radio pulsar 
emission. It was suggested that these phenomena could be due to the interaction between the neutron 
star magnetosphere and the surrounding debris disk. The pattern of pulsar emission depends on whether 
the disk can penetrate the light cylinder and efficiently quench the processes of particle production 
and acceleration inside the magnetospheric gap. A precessing disk may naturally account for the 
switch-on/off behavior in PSR B1931+24.

\subsection{Is PSR B0656+14 a very nearby RRAT source?}\smallskip
{\bf P.~Weltevrede et al.: } The recently discovered RRAT sources are characterized by very bright
radio bursts which, while being periodically related, occur infrequently. Bursts with the same
characteristics for the known pulsar B0656+14 were found. These bursts represent pulses from the 
bright end of an extended smooth pulse-energy distribution and were shown to be unlike giant pulses, 
giant micro-pulses or the pulses of normal pulsars. The extreme peak-fluxes of the brightest of 
these pulses indicates that PSR B0656+14, were it not so near, could only have been discovered 
as an RRAT source. Longer observations of the RRATs may reveal that they, like PSR B0656+14, 
emit weaker emission in addition to the bursts.

The emission of PSR B0656+14 can be characterized by two separate populations of pulses: bright pulses
have a narrow ``spiky'' appearance consisting of short quasi-periodic bursts of emission with microstructure,
in contrast to the underlying weaker broad pulses. The spiky pulses tend to appear in clusters which arise
and dissipate over about 10 periods. It was demonstrated that the spiky emission builds a narrow and peaked profile,
whereas the weak emission produces a broad hump, which is largely responsible for the shoulders in the total
emission profiles at both high and low frequencies.

\begin{figure*}[t!!]
  \centerline{\psfig{figure=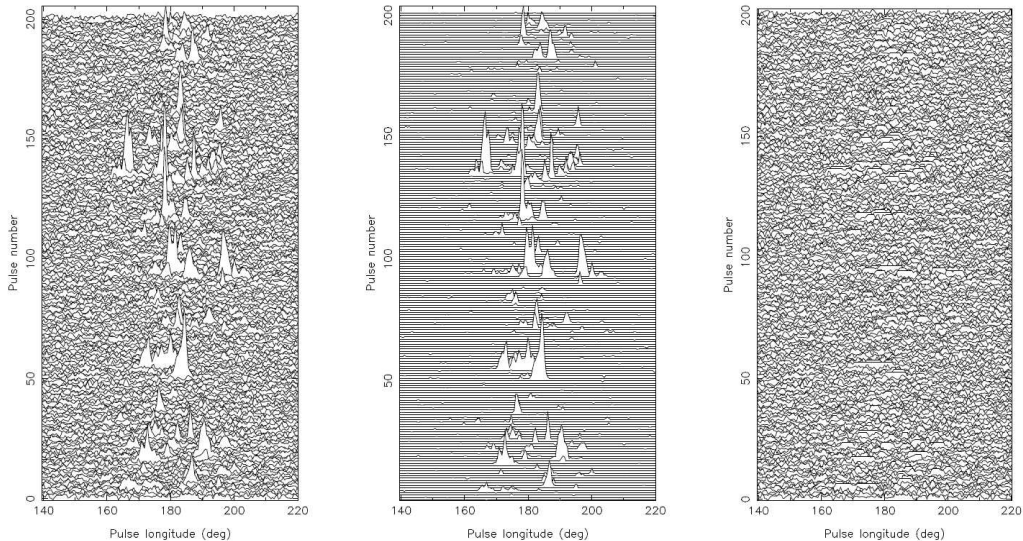,width=13.6cm,clip=}}
  \caption{A typical sequence of successive pulses (left panel) from PSR
  B0656+14. The same pulses are shown in the middle and right panel, but 
  there the emission is separated into the spiky and weak emission, 
  respectively. See poster 3.37 of Weltevrede et al. and astro-ph/0701189
  for further details.}
  \end{figure*}

\subsection{Glitch observations in slow pulsars}\smallskip
{\bf G.H.~Janssen, B.W.~Stappers: } An analysis of 5.5 years of timing observations of 7 ``slowly'' rotating 
radio pulsars, made with the Westerbork Synthesis Radio Telescope was presented. The improved timing solutions 
were presented and 30, mostly small, new glitches were found. Particularly interesting were the results on 
PSR J1814-1744, which is one of the pulsars with similar rotation parameters and magnetic field strength to the
anomalous X-ray pulsars (AXPs). Although the high-B radio pulsars don't show X-ray emission, and no radio 
emission is detected for AXPs, the roughly similar glitch parameters provide another tool to compare these 
classes of neutron stars. Furthermore, it was possible to detect glitches one to two orders of magnitude 
smaller than before, for example in our well-sampled observations of PSR B0355+54.  The total number of 
known glitches in PSR B1737-30 was doubled, and improved statistics on glitch sizes for this pulsar 
individually and pulsars in general were obtained.  No significant variations in dispersion measure for 
PSRs B1951+32 and B2224+65, two pulsars located in high-density surroundings, were detected. The effect 
of small glitches on timing noise was discussed. It was shown that it is possible to resolve timing-noise 
looking structures in the residuals of PSR B1951+32 by using a set of small glitches.

 \begin{figure*}
  \centerline{\psfig{figure=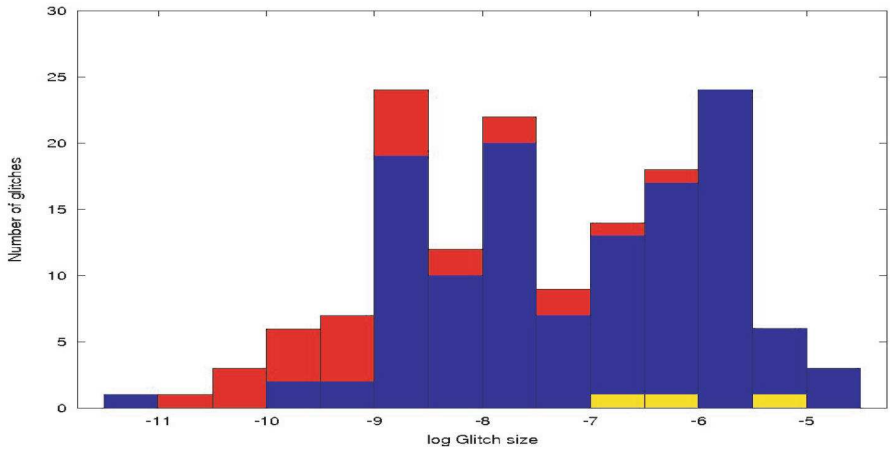,width=12cm,clip=}}
  \caption{Glitch size histogram showing all known glitches. Glitches found
  in this study are shown in red, AXP glitches (overlapping) in yellow. 
  See the poster 3.38 of Janssen \& Stappers and A\&A 2006, 457, 611.}
  \end{figure*}

\subsection{Mode coupling in pulsar magnetospheres due to plasma gradients perpendicular to the magnetic field}\smallskip 
{\bf A.C.~Judge: } Conventional ideas regarding plasma instabilities suggest that the
polarization of pulsar radio emission should be dominated by that corresponding to the fastest growing mode.
The presence of two distinct polarizations, indicating emission in two distinct modes, is, however, almost
ubiquitous in observations of these objects. In order to reconcile the basic theory with the observations
it has been proposed that energy is exchanged between the natural modes of the plasma as the radiation
propagates through the magnetosphere of the pulsar.

The basic theory of mode-coupling in stratified media has already been developed in work relating 
to wave propagation in the ionosphere and the solar corona. This formalism was applied here to a 
relativistically streaming plasma and gradients in the plasma perpendicular to the direction of the 
local magnetic field were investigated as a possible mechanism for effective mode-coupling in a 
pulsar magnetosphere.

\subsection{Software aspects of PuMa-II}\smallskip
{\bf R.~Karuppusamy, B.~Stappers: } The Pulsar Machine II (PuMa-II) is a state of the art pulsar machine,
installed at the Westerbork Synthesis Radio Telescope (WSRT), in December 2005. PuMa-II is a flexible
instrument and is designed around an ensemble of 44 high-performance computers running the Linux operating
system.  Much of the flexibility of PuMa-II comes from the software that is being developed for this
instrument. The radio signals reaching the telescope undergo several stages of electronic and software
processing before a scientifically useful data product is generated. The electronic processing of signals
includes the usual RF to IF conversion, analog to digital conversion and telescope dependent electronic
digital delay compensation that happen in the signal chain of WSRT. Within PuMa-II, this data is acquired,
stored and suitably processed. In this poster various aspects of PuMa-II software was presented and 
its pulsar signal processing capabilities were illustrated.

\subsection{High time resolution low-frequency pulsar studies}\smallskip
{\bf B.W.~Stappers: } Low frequency observations of radio pulsars have, to a certain extent, fallen out 
of favor in recent times. This is despite exciting and interesting work in Russia, Ukraine and India. 
The move to higher frequencies has mainly been due to the deleterious effects of the interstellar medium. 
However, with the increased availability of baseband recording and coherent de-dispersion techniques and 
new facilities such as the LFFEs at the Westerbork Synthesis Radio Telescope (WSRT) and in the future 
LOFAR/LWA, the interest in observations at frequencies below 300 MHz is growing again. Some exciting 
results on single pulse studies from observations at the WSRT (115-180 MHz) were presented here. These 
include the first full polarization observations of Crab giant pulses at these frequencies. The prospects 
for pulsar research with LOFAR. LOFAR will have unprecedented collecting area and bandwidth at frequencies 
below 220 MHz, allowing for a wide range of pulsar studies, in particular in emission physics, were also 
presented. The results from simulations which show that an all-sky survey with LOFAR could be expected 
to find up to 1500 new pulsars were shown. This survey would provide significant constraints on the 
low-end of the pulsar luminosity distribution which has important consequences for the total pulsar 
population. It was argued that LOFAR could detect pulsars in nearby galaxies.

\subsection{The 8gr8 Cygnus survey for New pulsars and RRATs}\smallskip
{\bf E. Rubio-Herrera et al.: } A survey to search for new pulsars and the recently found Rotating 
RAdio Transients (RRATs) in the Cygnus OB complex was currently undertaken. The survey uses the 
Westerbork Synthesis Radio Telescope in a unique mode which gives it the best sensitivity of any
low-frequency wide-area survey. Few new pulsars were found so far. The program of using routines 
for the detection of RRATs. Some initial results on the new pulsars and possible transients were 
presented. It is expected to find a few tens of new pulsars and a similar number of RRATs. The latter
discoveries should help to improve the  knowledge about the population and properties of the  poorly 
known objects as well as provide an improved knowledge of the number of young pulsars associated 
with the OB complexes in the Cygnus region.

\subsection{Pulsar coherent de-dispersion observation at Urumqi Observatory}\smallskip
{\bf A.~Yishamuding: } Based on a Mark5A VLBI backend and a four node cluster, pulsar off-line 
coherent de-dispersion observations have been conducted by using Urumqi 25m telescope. The 
observing system was described and the initial results were presented in this paper.

\subsection{X-ray monitoring of the pulsar PSR B1259-63}\smallskip
{\bf H.H.~Huang, W.~Becker: } PSR B1259-63, a rotation-powered radio pulsar with a $\sim 48$ millisecond 
period, is in a highly eccentric ($e \sim 0.87$) 3.4 year orbit around a massive Be star SS 2883. The results 
of the XMM-Newton observations performed between 2001 and 2004 were summarized. Combining the XMM-Newton 
observations with the previous results from ASCA, it was found that the best-fit power-law models in 
$1.0-10.0$ keV energy band show long term variations in the photon indices from $\sim 1.11$ to $\sim 1.95$. 
The X-ray flux was observed to increase by a factor of $> 10$ from apastron near to periastron. No X-ray 
pulsations at the pulsar's spin period were found in any observations so far. A model invoking the interaction 
between the pulsar and the stellar wind was likely to explain the observed orbital phase-dependent 
time variability in the X-ray flux and spectrum.

\subsection{XMM-Newton observation of PSR B1957+20}\smallskip
{\bf H.H.~Huang, W.~Becker: } The "Black Widow pulsar", PSR B1957+20, is a millisecond pulsar which is 
in a 9.16 - hour binary system. H$_\alpha$ bow-shock nebula created by the interaction between the 
relativistic wind of the pulsar and the surrounding ISM and ablation of the low-mass companion star by 
the pulsar wind were observed. 30 ksec observation of PSR B1957+20, using the EPIC-MOS detector on-board 
the XMM-Newton observatory, were reported. The X-ray diffused emissions detected from this source was found
to be consistent with the results derived from Chandra observations. The spectrum of the nebular emission 
was modeled with a single power law spectrum of photon index $2.1^{+0.4}_{-0.3}$. This extended emission
generated by accelerated particles in the post shock flow was considered to explain this result. For the 
first time, a significant X-ray flux modulation near to the pulsar's radio eclipse was detected.

 \begin{figure*}[h!!]
  \centerline{\psfig{figure=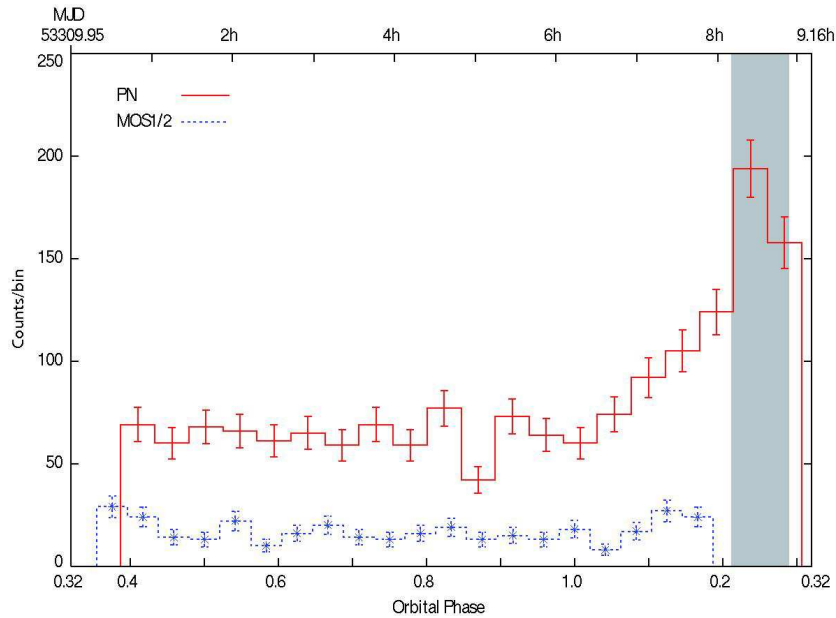,width=11cm,clip=}}
  \caption{X-ray emission from PSR B1957+20 within $0.3-3.0$ keV as function
   of the pulsar's orbital phase ($\phi$). One complete orbital period of 
   this system is mapped with the starting point $MJD=53309.95$, i.e. 
   $\phi = 0.32$. $\phi = 1.0$ corresponds to the ascending node of the 
   pulsar orbit. The upper curve was obtained from the XMM-Newton EPIC-PN 
   (background level at 65 cts/bin). The lower lightcurve is obtained from 
   the MOS1/2 data. The gray strip between the orbital angle $0.21 - 0.29$ 
   indicates the eclipse of the pulsar. Phase bins with zero
   counts correspond to phase angles not covered in the observation. Image
   from the poster by Huang \& Becker. See also http://arxiv.org/abs/astro-ph/0701611.}
  \end{figure*}

\subsection{Optical observations of binary millisecond X-ray pulsars in quiescence}\smallskip
{\bf P.~Callanan et al.:} The discovery of accreting binary millisecond pulsars finally provided 
firm confirmation of the link between bright accreting Low Mass X-ray Binaries and millisecond pulsars. 
Little is known about their optical properties in quiescence, however. Here the optical observations 
of SAX J1808.4-3658 and IGR J00291+5934 in quiescence were presented, and comparison of them to other 
quiescent X-ray transients was made.

\subsection{X-Ray studies of the central compact objects in Puppis-A \& RX J0852.0-4622}\smallskip
{\bf C.Y.~Hui, W.~Becker: } The Supernova remnants (SNRs) Puppis-A and RX J0852.0-4622 (Vela-Junior) 
are located along the line of sight towards the outer rim of the Vela SNR. Central compact objects 
(CCOs) were discovered in each of them. Both CCOs are thought to be the compact stellar remnants
formed in core-collapsed supernova explosions. Nevertheless, the emission properties observed from 
these sources were found to be completely different from what is observed in other young canonical 
neutron stars. Based on observations with the X-ray observatories Chandra and XMM-Newton, the most recent
results from a detailed spectro-imaging and timing analysis of these two enigmatic sources were presented.

  \begin{figure*}[h!!]
  \centerline{\psfig{figure=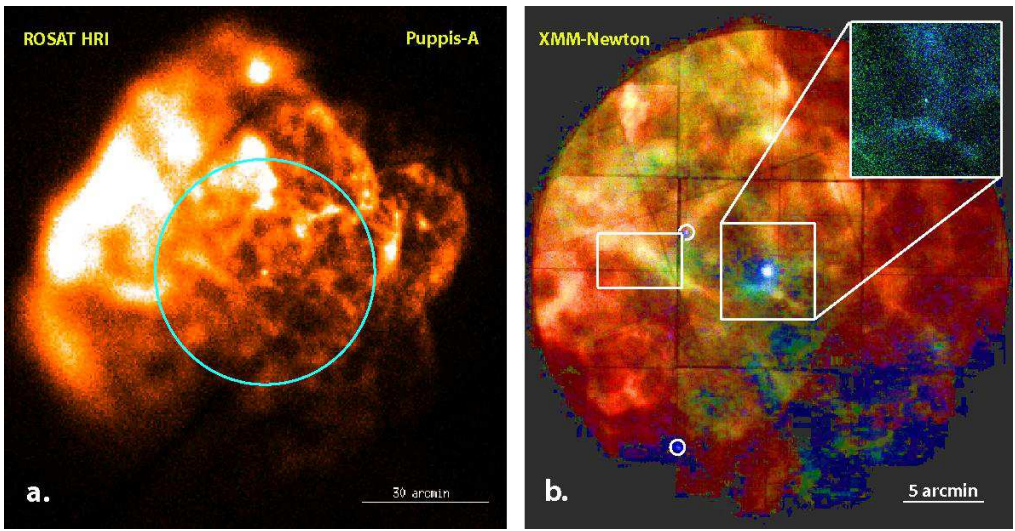,width=13.4cm,clip=}}
  \caption[]{\small {\bf a.}~Composite ROSAT HRI image of the Puppis-A
   supernova remnant. The blue ring indicates the 30 arcmin central 
   region which has been observed by XMM-Newton. {\bf b.}~XMM-Newton 
   MOS1/2 false color image of the central region of Puppis-A 
   (red: $0.3 - 0.75$ keV, green: $0.75 - 2$ keV and blue: $2-10$ keV). 
   The central source is the CCO RX~J0822$-$4300. The inset shows the 
   squared region as observed by the Chandra HRC-I. See poster 3.47 of 
   Hui \& Becker and A\&A 2006, 454, 543 for further details.}
  \end{figure*}

\subsection{Probing the proper motion of the central compact object in Puppis-A}\smallskip
{\bf C.Y.~Hui, W.~Becker: } Using two observations taken with the High Resolution Camera (HRC-I) 
aboard the Chandra X-ray satellite, we have examined the central compact object RX J0822-4300 
for a possible proper motion. The position of RX J0822-4300 was found to be different by 
$0.574\pm 0.184$ arcsec, implying a proper motion of $107.49\pm 34.46$ mas/yr with a position 
angle of $241\pm 24$ deg. For a distance of 2.2 kpc, this proper motion is equivalent to a 
recoil velocity of $1121.79\pm 359.60$ km/s. Both the magnitude and the direction of the 
proper motion are in agreement with the birth place of RX J0822$-$4300 being near to the 
optical expansion center of the supernova remnant. Although this is a promising indication 
of a fast moving compact object in a supernova remnant, the relative large error prevents 
any constraining conclusion.

 \begin{figure*}
  \centerline{\psfig{figure=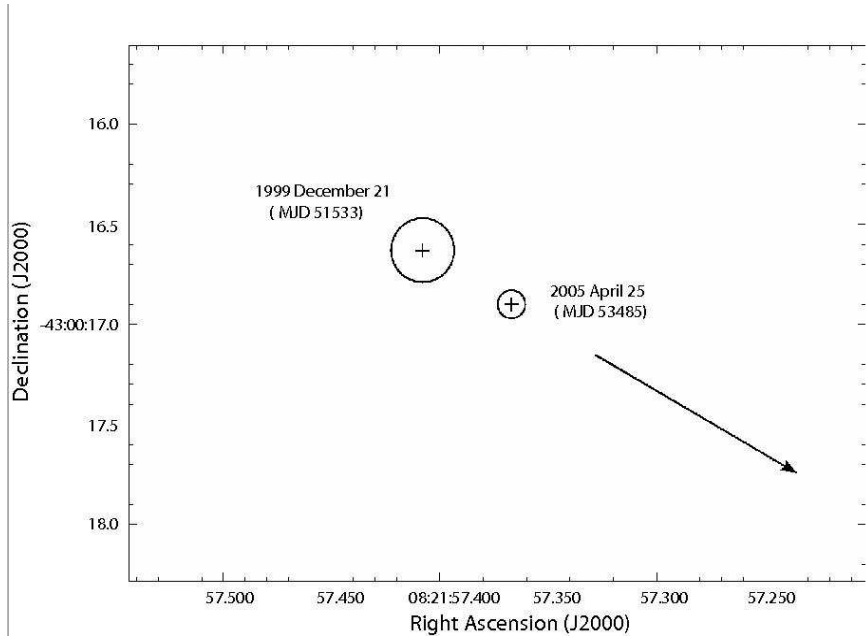,width=12cm,clip=}}
  \caption{The best-fitted X-ray positions of RX J0822-4300 at two different
  epochs separated by 1952 days are marked by crosses. The circles indicate
  the $1-\sigma$ error. The arrow shows the direction of proper motion
  inferred
  from both positions. Image from the poster 3.48 of Hui \& Becker. See also
  Hui \& Becker 2006, A\&A,457, 33.}
  \end{figure*}

\subsection{Exposing drifting subpulses from the slowest to the fastest pulsars}
{\bf J.~van Leeuwen: } Pulsar emission is surprisingly similar over a vast range of periods 
and magnetic fields: all the way from the 2-millisecond $10^8$ G recycled pulsars to the 
6-second $10^{14}$ G magnetar-like regular pulsars. It was investigated how the curious 
instabilities called 'drifting subpulses' can discern between different mechanisms 
for pulsar emission.

\subsection{Pulsar research with LOFAR, the first next-generation radio telescope}
{\bf J.~van Leeuwen, B.~Stappers: } LOFAR is a low-frequency radio telescope of revolutionary 
design that is currently being constructed and will become operational in 2007. In stark contrast 
to radio dishes, LOFAR is the first telescope that relies on a central supercomputer to combine 
the signals of ten thousand individual dipoles to form several extremely sensitive, independently 
steerable beams on the sky. It was discussed how LOFAR opens up a new frequency window with 
unprecedented sensitivity and why LOFAR will have considerable impact on radio pulsar research.

 \begin{figure*}
  \centerline{\psfig{figure=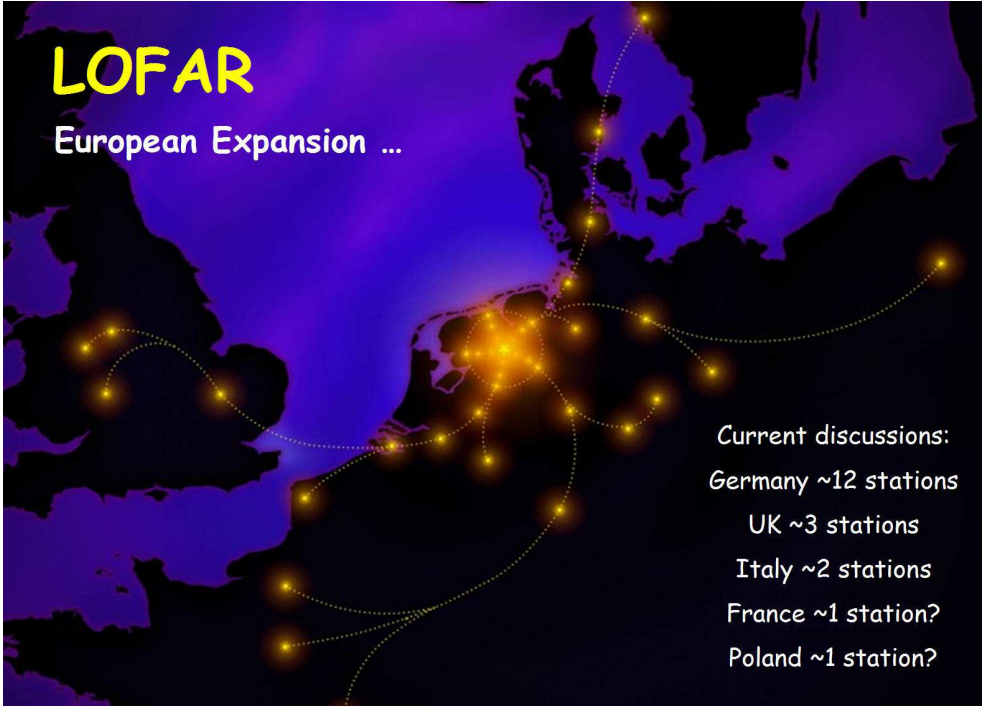,width=13cm,clip=}}
  \caption{Chart of the distribution of radio antennas as planed in the
  LOFAR project as presented in the talk by M.~Kramer and in the poster 3.50 by
  J.~van Leeuwen and B.~Stappers.}
  \end{figure*}

\subsection{Non-dipolar surface magnetic field of neutron stars: General approach and observational consequences}\smallskip
{\bf G.I.~Melikidze, A.~Szary, J.~Gil: } It is widely accepted that the magnetic field 
structure near the surface of neutron stars may significantly differ from the star centered 
global dipole structure. Due to flux conservation of the open magnetic field lines, strong 
non-dipolar surface field results in significant shrinking of the canonical polar cap, in 
general. We have modeled different possible configurations and found out that for some 
configurations the pair creation is possible not only along the open field lines, but also 
in the region of closed field lines. Therefore, in this case, we can naturally explain 
some peculiarities of pulsar activities, such as unusual thermal x-ray emission, reversible 
radio emission and rotating radio transients.

The pairs created along the closed field lines can easily reach the stellar surface near 
the polar cap at the opposite side of the neutron star and heat the surface area that can 
even exceed that of the canonical polar cap. Both smaller (often) and larger (rarely) 
bolometric surface areas of the hot polar cap are observed.

In the frame of this model, we can easily realize the configuration, which allows the pair 
creation near both polar caps (along the same field). In this case, two streams of the pair 
plasma penetrate each other creating a favorable condition for the two-stream instability 
to be developed. Such a process can lead to the radio emission generation, either in 
quasi-stationary or stochastic process. Consequently, either quasi-stationary reversible 
radiation, or stochastic emission of the transients can be observed.

\subsection{Glitches in the Vela pulsar}\smallskip
{\bf S.~Buchner, C.~Flanagan:} The Vela pulsar undergoes occasional sudden spin-ups in rotational 
frequency. The recovery from these glitches provides insight into the internal structure of the 
neutron star. The HartRAO was used to monitor Vela since 1984 and eight large glitches were
observed. These data were presented in this paper.

\subsection{Optical polarization of the Crab pulsar with 10 $\mu s$ time resolution}\smallskip
{\bf A.~Slowikowska, G.~Kanbach, A.~Stefanescu: } The Crab nebula and pulsar were observed for 
about 25 hours with the high-speed photo-polarimeter OPTIMA  in November 2003 at the Nordic 
Optical Telescope, La Palma. The instrument's sensitivity (white light) extends from about 
450nm to 950nm and reaches about 60 \%. Linear polarization is measured with a continuously 
rotating polaroid filter which modulates the incoming radiation. The astronomical target is 
viewed through the polaroid and imaged onto a hexagonal bundle of optical fibers which are 
coupled to single photon APD counters. The spacing and size of the fibers at NOT corresponds 
to about 2 arcsec. GPS based time tagging of single photons with 4 microsec resolution, 
together with the instantaneous determination of the angular position of the rotating 
polaroid filter, allowed to measure the phase dependent linear polarization state of the 
pulsar and the surrounding nebula simultaneously.

\begin{figure*}[h!!]
  \centerline{\psfig{figure=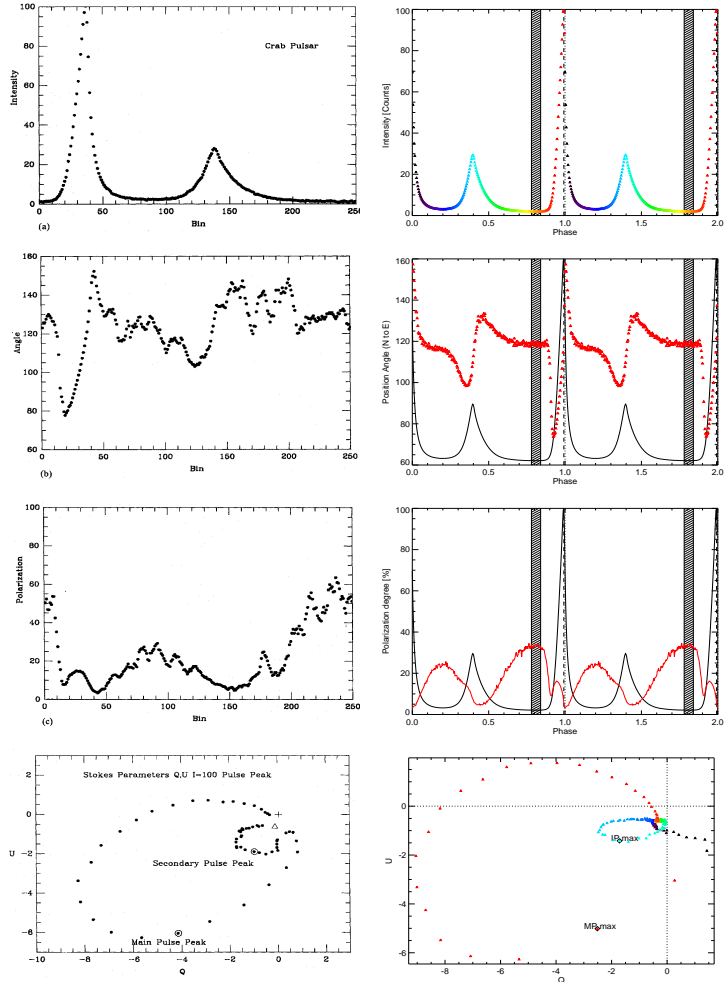,width=9.5cm,clip=}}
  \caption{Comparison of the optical polarization of the Crab pulsar 
   obtained by Smith et al.~1988 (left) and Slowikowska, Kanbach,
   and Stefanescu (right). From top to bottom: the intensity, position 
   angle, and polarization degree are plotted as a function of the 
   pulsar phase, while the Stokes parameters Q, U are plotted as
   a vector diagram. There are 
   250 bins per cycle in both cases, the only difference is that for 
   clarity two periods are shown in the right chart. As a DC component 
   Smith et al.~took 50 out of 250 bins, whereas in the right chart 
   only 7\% of the rotational period was taken. Figure and caption 
   from the poster 3.53 by Slowikowska et al.}
  \end{figure*}

The Crab pulsar and its net optical polarization were determined at all
phases of rotation with extremely high statistical accuracy. On time 
scales of a few tens of microseconds significant details of the 
polarization of the main emission peak became visible.

\end{document}